\documentclass[final,11pt,5p,times,twocolumn,sort&compress]{elsarticle}
\usepackage{amsmath}
\usepackage{amssymb}
\usepackage{multirow}
\usepackage{bm}
\usepackage{textcomp}
\usepackage{gensymb} 
\usepackage{esint} 
\usepackage[unicode=true,colorlinks=true,linkcolor=blue,citecolor=blue,urlcolor= blue]{hyperref} 
\usepackage[english]{babel}
\usepackage[latin1]{inputenc}
\usepackage[T1]{fontenc}
\usepackage{graphicx}
\usepackage{adjustbox}
\usepackage{array}
\begin{document}
\begin{frontmatter}
\title{\textit{ab initio} investigation of phosphorus and hydrogen co-segregation and embrittlement in
\texorpdfstring{$\alpha$}--Fe twin boundaries}

\address[EMSE]{Mines Saint-Etienne, Univ Lyon, CNRS, UMR 5307 LGF, Centre SMS, F-42023 Saint-Etienne France}
\address[SRMP]{DEN-Service de Recherches de M\'etallurgie Physique, CEA, Universit\'e Paris-Saclay, F-91191 Gif-sur-Yvette, France}
\cortext[cor1]{Corresponding author: thomas.schuler@cea.fr}
\author[EMSE,SRMP]{Thomas Schuler \corref{cor1}}
\author[EMSE]{Fr\'ed\'eric Christien}
\author[EMSE]{Patrick Ganster}
\author[EMSE]{Krzysztof Wolski}

\date{\today}

\begin{abstract}
We propose a new statistical physics model to study equilibrium solute segregation at grain boundaries and the resulting embrittlement effect. This low-temperature expansion model is general and efficient, and its parameters can be obtained from atomistic calculations. It is possible to take into account multiple species, multiple segregation sites with different segregation free energies, account for configurational entropy, grain radius and site competition between solutes.
As an example, the model is then applied to the study of phosphorus and hydrogen co-segregation at $\Sigma3$ $109.5\degree\left[0\bar{1}1\right]\left\{ 111\right\}$ twin boundaries in $\alpha$-Fe, using energetic parameters from density-functional theory calculations. We show that P--H interactions may lead to increased P segregation at grain boundaries and cause additional embrittlement compared to the case where P and H are considered separately.
\end{abstract}

\end{frontmatter}

\section{Introduction}

Grain boundaries (GBs) are of prime importance in the study of materials because they form an interconnected network that spreads over the whole material and directly affects its mechanical properties. Hence, the study of GBs has drawn a lot of attention for more than 60 years \citep{McLean1957, Lejcek2010, Priester2013}. Nowadays, a certain level of understanding of GB thermodynamics and kinetics is reached, such that microstructure optimization via GB engineering approaches are emerging \citep{Priester2013, RAABE2014}. One of the most studied feature of GBs is their interaction with solutes from the bulk material, because segregation phenomena lead to large and very localized chemical inhomogeneities \citep{Seah1973}, and atomic bonding at GB planes are found very sensitive to the presence of solutes \citep{Wu1994}.

There are many difficulties in investigating GB segregation experimentally: ideally, we would need a technique that is able to probe very localized regions of the material while making sure that the chemical distribution is not altered by the measurement, and this technique should be able to give accurate chemical information--sometimes for very low concentrations--and crystallographic information, as well as being fast enough to produce sufficient statistics over a large number of GBs. Consequently, the study of GB segregation usually resorts to a combination of experimental techniques \citep{Herbig2014,Christien2019} and results depend on the annealing temperature, bulk chemical composition, chemical environment and GB structure. The task becomes even more complicated when studying co-segregation effects, \textit{i.e.} the simultaneous segregation of at least two chemical species which occurs all the time in real-life materials. There are several mechanisms at work: solute-GB interaction, solute-solute interaction at the GB and in the bulk, site competition, concentration ratio between solutes and kinetic properties of each solute \citep{Guttmann1983}. For instance in Fe-P-C alloys, it was noted that intergranular P segregation decreases with increasing C content but the reason why this happens is not clear \citep{Erhart1981,Christien2019}. If Cr is added to this alloy it seems that P segregation increases only when C concentration is above a certain level \citep{Erhart1981}.

The pioneering work of Wu \textit{et al.} \citep{Wu1994} on $\Sigma 3[1\bar{1}0](111)$ GB in $\alpha$-Fe showed that \textit{ab initio} methods were able to provide some insight on solute properties at GBs, as the authors were able to explain the strengthening by B additions and the embrittlement by P additions in terms of electronic structure at the GB and at free surfaces. These methods are fast-growing, and computation capabilities have increased a lot such that \textit{ab initio} studies on GBs are now standard. For instance, it is now possible to study the segregation of various solutes (B,C,N,O,P) on various GBs ($\Sigma 3$ and $\Sigma 5$) at various concentrations (1.0, 0.5 and 0.25 mono-layers) in a systematic manner \citep{Wachowicz2011}. Moreover, there are some efforts to regroup the large collection of \textit{ab initio} data computed over the years and extract some trends as a function of solute properties \citep{Gibson2015, Gibson2016, Lejek2017}. Most of this huge amount of data is restricted to the segregation of one type of solute, and most of the time one type of segregation site. Yet, there are some studies focusing on more complex segregation phenomena, e.g.  strong co-segregation of Mn-C\citep{Wicaksono2017} or site exclusion between N and C\citep{MATSUMOTO2011}. These effects were shown using \textit{ab initio} calculations but a comprehensive analysis using statistical physics tool is still missing. 

Of course these \textit{ab initio} methods have their drawbacks, mostly related to the complexity of properly taking into account temperature effects, and the limited number of atoms that can be included in a simulation box. Regarding the latter, a question was raised about the reliability of \textit{ab initio} calculations for solutes with small solubility limits \citep{Lejcek2013, Lejcek2018-OpenQuestion}. To shorten the introduction, we briefly comment on this question in \ref{sec:Appendix0}.

A review of the various statistical models that have been developed to study GB segregation can be found in Refs. \citep{Hondros1977, Lejcek2010}. The most widely used model is the simplest one, the Langmuir-McLean isotherm \citep{Langmuir1918, McLean1957}, which was extended to multi-component systems by Guttmann \citep{Guttmann1975,Guttmann1983}. The limitation of these approaches is that they are derived from mean-field arguments. Even though there are some semi-empirical procedures to derive the mean-field interaction parameters \citep{Lejcek2018}, the mean-field model cannot always render the full complexity of the segregation and the connection with \textit{ab initio} binding energies is not straightforward. There were some attempts to consider explicitly the short-range order in the first nearest-neighbor shell \citep{Lupis1967} but this model is not easy to generalize to more complicated case. A model for multi-site segregation was proposed \citep{NOWICKI1990}, but again the extension to multi-species segregation is not straightforward. The embrittlement potency of a solute is defined as the solute segregation energy difference between the GB and the two free surfaces obtained after fracture \citep{RiceWang1989}. As noted earlier \citep{Wang2016}, this thermodynamic definition is not valid for more complicated cases, namely multi-site segregation or co-segregation with competing embrittlement potencies and this data is now becoming available (\textit{e.g.} with molecular statics simulations\citep{Rajagopalan2013}). In a nutshell, it seems that nowadays we have access to a large number of atomic-scale data, but we are missing some efficient and general model to take full advantage of this data.

Let us now focus on the practical case of P and H co-segregation in $\alpha$-Fe. H \citep{Rogers1968} and P \citep{Spitzig1972} are both known to alter fracture toughness of steels with very small amounts of impurities. 
Aucouturier studied a Fe-P-H steels and was not able to tell if P simply reduces the critical H concentration for hydrogen-induced cracking, or if GB embrittlement was due to P-H interactions at the GB \citep{Aucouturier1982}. 
According to Kameda \textit{et al.} the effects of P and H are additive \citep{Kameda1983} meaning that P and H do not interact at all, or in other words, there is no co-segregation effect. According to Komazazki, P does not influence the H content of GBs but weakens the GB, which translates into increased H embrittlement \citep{Komazazki2003}. This was somehow confirmed by McMahon who emphasizes the dynamic effect of H and its low segregation enthalpy to GBs \citep{McMahon2004}. Solubilities of H and P are very low in these materials which complicates the precise determination of P-H co-segregation effects. It seems that modern modeling tools can shed light on the interaction between P and H at the GB. $\Sigma3$ $109.5\degree\left[0\bar{1}1\right]\left\{ 111\right\}$ twin boundaries are among the most common ones in lath martensite in low carbon steels \citep{Morito2006} and they are quite easily modeled with \textit{ab initio} methods because they are high-angle symmetric tilt GBs with a well-known structure \citep{Gao2009}. Also, due to the dense coincidence site lattice, the range of elastic deformation around the GB is expected to be small such that segregation occurs mainly in the grain boundary plane and we can reduce the size of the simulation cell and the number of configurations to explore.

In this paper, we study the co-segregation of P and H at $\Sigma3$ $109.5\degree\left[0\bar{1}1\right]\left\{ 111\right\}$ twin boundaries in $\alpha$-Fe. First, we introduce the modeling tools (Sec. \ref{methods}), namely density functional theory to compute binding and segregation energies, and low-temperature expansions. The latter provide a new type of statistical model to study grain boundary segregation and has the advantage of being general and straightforward to use in conjunction with \textit{ab initio} calculations. It can take into account competing segregations (fixed solute concentration distributed among bulk and various GBs) and also the effect of grain size. Then we present \textit{ab initio} calculations of P-H interactions in the bulk, at the GB and at a free surface (Sec. \ref{dft}). This data is finally used in the low-temperature expansion model to conclude about the effect of the simultaneous presence of H and P on segregation levels and embrittlement potency (Sec. \ref{results}). Finally, we discuss our modeling assumptions (Sec. \ref{discussion}).

\section{Methods \label{methods}}

\subsection{Density-functional theory\label{subsec:Density-functional-theory}}

Density functional theory (DFT) calculations were performed using the PWscf
package from the {\sc Quantum ESPRESSO} software suite \citep{Gianozzi1,Gianozzi2}.
The exchange and correlation energy is computed with the PBE generalized
gradient approximation \citep{Perdew1996}. Hydrogen $\left(1s^{1}\right)$,
phosphorus $\left(\left[\mathrm{Ne}\right]3s^{2}3p^{3}\right)$ and
iron $\left(\left[\mathrm{Ar}\right]4s^{2}3d^{6}\right)$ nuclei and
core electrons are modeled by a projector augmented wave potential
\citep{Kresse1999}, as generated by Dal Corso \citep{pseudo1,pseudo2}.
The kinetic energy cutoff for wave-functions is set to 870 eV, which
ensures energy convergence below 0.01 eV/atom. Calculations are performed
on 54, 76 and 144 atoms supercells for bulk, surface and GB
calculations, respectively, with a $4\times4\times4$, $5\times5\times2$
and $5\times5\times2$ gamma centered k-point mesh. First-order Methfessel-Paxton
smearing \citep{Methfessel1989} with an energy set to 0.27 eV ensures
accurate evaluation of forces for ionic relaxation. Binding energy
calculations are relaxed with a conjugate gradient method, until the
force on each ion is below 0.025 eV/\AA. For GB calculations,
supercell relaxation is performed along the direction perpendicular
to the GB plane, with a pressure threshold equal to 0.1
kbar. For surface calculations, a void equal to the slab thickness
was used. Using a void that is half or twice as thick gives identical
surface energy values within less than 1 meV. All calculations are
spin-polarized with collinear magnetization to reproduce iron ferromagnetic
state. With these settings the lattice parameter for body-centered
cubic Fe was found equal to $a=2.826$ \AA, in good agreement with
experimental measurements $a=2.87$ \AA  \citep{Kittel} and other
DFT calculations $a=2.832$ \AA  \citep{Fellinger2017}. We computed
the elastic constants using three volume-conserving strains of a Fe
primitive cell \citep{Mehl1994,ElastConst} and found the following
values $C_{11}=305$ GPa, $C_{12}=146$ GPa and $C_{44}=84$ GPa,
in fair agreement with other DFT calculations ($C_{11}=278$ GPa,
$C_{12}=148$ GPa and $C_{44}=98$ GPa \citep{Fellinger2017}). Zero-point
energies are included for all supercells containing hydrogen atoms,
as this correction is not negligible for hydrogen atoms \citep{JiangCarter,Hayward2013}.
We could not afford the rigorous calculation of the force constant
matrix where all atoms in the simulation cell are displaced so we
limit ourselves to displacing the hydrogen atom only, by 0.02 \AA{} in six directions. This approximation is justified by the large mass difference between hydrogen and iron atoms \citep{JiangCarter, COUNTS2010}. The zero-point energy correction at $T=0$ K is then expressed as:

\begin{equation}
E_{zpe}^{H}=\dfrac{h}{2}\sum_{i}\nu_i=\frac{\hbar}{2}\sum_{i}\sqrt{\dfrac{k_{i}}{m_{H}}},\label{eq:zpe}
\end{equation}
where $\hbar$ is the reduced Planck constant, $m_{H}$ the hydrogen atom mass and $k_{i}$ are the eigenvalues of the force-constant matrix.
For any temperature, the H atom vibration contribution to the total energy of the system is \citep{Schober1995, Fultz2010}:
\begin{equation}
    F_{vib}^{H} = k_{B}T\sum_{i}\ln\left(2\sinh\left(\dfrac{h\nu_{i}}{2k_{B}T}\right)\right),\label{eq:Fvib} 
\end{equation}
and Eq. \ref{eq:Fvib} reduces to Eq. \ref{eq:zpe} at $T=0$ K. In the  calculations performed in Sec. \ref{results} Eq. \ref{eq:Fvib} is used to compute the vibrational contribution to the free energy of the system.

In this paper, a positive binding energy reflects a configuration
which is more stable than all solutes being isolated in their reference
states. The reference states are chosen as the most stable P and H
sites in bulk Fe, substitutional and interstitial tetrahedral, respectively.
The binding energy $E^{b}$ between P and H in a bulk system is defined
as:

\begin{align}
E^{b}\left(\mathrm{PH}\right) & =E\left[53\mathrm{Fe}+\mathrm{P}_{sub}\right]+E\left[54\mathrm{Fe}+\mathrm{H}_{tet}\right]\nonumber \\
 & -E\left[54\mathrm{Fe}\right]-E\left[53\mathrm{Fe}+\mathrm{P}+\mathrm{H}\right],\label{eq:Eb}
\end{align}
where $E\left[\alpha\right]$ is the energy of a supercell containing
$\alpha$ atoms, computed using DFT. In order to keep the same references
to compute binding energies at surfaces or GBs, the following
definition is used:

\begin{align}
E^{b}\left(\mathrm{PH}\right) & =E\left[53\mathrm{Fe}+\mathrm{P}_{sub}\right]+E\left[54\mathrm{Fe}+\mathrm{H}_{tet}\right]+E\left[nFe,\:d\right]\nonumber \\
 & -2E\left[54\mathrm{Fe}\right]-E\left[\left(n-1\right)\mathrm{Fe}+\mathrm{P}+\mathrm{H},\:d\right],\label{eq:Eb_defect}
\end{align}
where $E\left[\alpha,\:d\right]$ is the energy of a supercell containing
$\alpha$ atoms and an extended defect $d$ which in our case is either
a surface or a GB. With this definition (Eq. \ref{eq:Eb_defect}), the binding energy at an interface contains the P-H interaction at the interface, as well as the P-interface and H-interface interactions.

\subsection{Low-temperature expansion\label{subsec:Low-temperature-expansion}}

The segregation of chemical species to interfaces is generally quantified
using the McLean model \citep{McLean1957,Hondros1977,Lejcek2014}.
Unfortunately, this model is limited to one segregated species at
one specific segregation site. A generalization of this model to multi-component
framework was derived by Guttmann \citep{Guttmann1975,Guttmann1983},
based on a mean-field approach. The mean-field does not allow to take
into account configurations with specific short-range order and segregation
energies. In this paper, the concentrations of segregated elements are computed
using a low-temperature expansion formalism (LTE) \citep{Ducastelle1991,LeBouarLTE,Clouet2007,SchulerPRL}.
This formalism is easier to generalize to several chemical
species and a variety of segregation sites and configurations. It
is also able to treat systems in both canonical (fixed concentration)
and grand-canonical (fixed chemical potential) ensembles. Finally,
it is in principle straightforward\textendash even though practically
challenging\textendash to take into account the overlap and interaction
between segregated species at large coverage. In \ref{sec:Appendix},
we show that the LTE formalism is rigorously equivalent to McLean's
model in the limit of non-interacting particles having the same segregation
energy.

For the system of interest, the grand potential $\mathcal{A}$ is
related to the grand-canonical partition function $\left(Z\right)$
with the standard relation:

\begin{align}
\mathcal{A} & =-k_{B}T\ln\left(Z\right)\nonumber \\
 & =-k_{B}T\ln\left(\sum_{i}G_{i}\exp\left(\dfrac{-E_{i}+\sum_{\alpha}n_{i}^{\alpha}\mu_{\alpha}}{k_{B}T}\right)\right),\label{eq:LTE1}
\end{align}
where $k_{B}$ is the Boltzmann constant, $T$ is the absolute temperature
and the sum runs over micro-states $i$ characterized by energy $E_{i}$
and containing $n_{i}^{\alpha}$ atoms of chemical species $\alpha$.
$G_{i}$ corresponds to the degeneracy of micro-state $i$ and $\mu_{\alpha}$
is the chemical potential associated with species $\alpha$. Note
that for substitutional solutes, $\mu_{\alpha}$ denotes the difference
$\mu_{\alpha}-\mu_{m}$ because it is necessary to remove a matrix
atom $m$ from the system in order to insert the solute while keeping
the total number of sites constant. We now define a reference state
with energy $E_{0}$ and containing $\{n_{0}^{\alpha}\}$ atoms for each species $\alpha$. In our case,
the reference system is the ideal interface without any atoms other than matrix (Fe) atoms. The addition of solutes to the system
will be treated as excitations with respect to this reference state.
\begin{align}
 & \mathcal{A}=\mathcal{A}_{0}-\nonumber \\
 & k_{B}T\ln\left(1+\sum_{i\neq0}G_{i}\exp\left(\dfrac{E_{0}-E_{i}+\sum_{\alpha}\delta n_{i}^{\alpha}\mu_{\alpha}}{k_{B}T}\right)\right),\label{eq:LTE2}
\end{align}
where $\mathcal{A}_{0}=E_{0}-\sum_{\alpha}n_{0}^{\alpha}\mu_{\alpha}$ and $\delta n_{i}^{\alpha}=n_{i}^{\alpha}-n_{0}^{\alpha}$. The logarithm function is then Taylor expanded to infinite order and the linked-cluster theorem shows that all terms that are non-linear in the number of sites $N$ cancel out \citep{Ducastelle1991}. The chemical potential is the energy required to take a given atom from a reservoir
to a reference site in the system. Taking these reference sites 
identical to those in Eqs. \ref{eq:Eb} and \ref{eq:Eb_defect}, $-\left(E_{i}-E_{0}\right)=E_{i}^{b}$,
the binding energy of atoms in configuration $i$. Also, because all
remaining excited states are proportional to $N$, we can define $g_{i}=\lim_{N\rightarrow0}\left[G_{i}/N\right]$
as the degeneracy per site.

\begin{equation}
\frac{\mathcal{A}}{N}=\frac{\mathcal{A}_{0}}{N}-k_{B}T\sum_{i\neq0}g_{i}\exp\left(\dfrac{E_{i}^{b}+\sum_{\alpha}\delta n_{i}^{\alpha}\mu_{\alpha}}{k_{B}T}\right),\label{eq:LTE3}
\end{equation}
The concentration of atoms of species $\alpha$ is computed as the
derivative of $\mathcal{A}/N$ with respect to $\mu_{\alpha}$:

\begin{equation}
\left[\alpha\right]=-\frac{1}{N}\frac{\partial\mathcal{A}}{\partial\mu_{\alpha}}=\dfrac{n_{0}^{\alpha}}{N}+\sum_{i\neq0}\delta n_{i}^{\alpha}g_{i}\exp\left(\dfrac{E_{i}^{b}}{k_{B}T}\right)\prod_{\alpha}X_{\alpha}^{\delta n_{i}^{\alpha}}.\label{eq:LTE4}
\end{equation}
A similar equation must be written for each species $\alpha$, which
gives a system of coupled polynomial equations whose unknowns are
quantities $X_{\alpha}=\exp\left(\mu_{\alpha}/k_{B}T\right)$. Either
these variables $X_{\alpha}$ are known (grand-canonical ensemble)
and it is straightforward to compute the total solute concentrations,
or nominal concentrations $\left[\alpha\right]$ are known and one needs to
solve the system of equations. The quantity $g_{i}\exp\left(E_{i}^{b}/k_{B}T\right)\prod_{\alpha}X_{\alpha}^{\delta n_{i}^{\alpha}}$ represents the concentration per site of atoms in the specific configuration of micro-state $i$. 

The sum over micro-states $i$ in Eqs. \ref{eq:LTE3} and \ref{eq:LTE4}
is theoretically infinite but states with a low energy value (binding
energy plus chemical potential contribution) are negligible. In other
words, we only need to keep in the series states of low excitation
energies, which are usually the states with a small number of solute atoms that are different from those of the reference states.

There are two equivalent ways to deal with Eqs. \ref{eq:LTE3} and
\ref{eq:LTE4} in our case: either consider a single system containing
bulk and interface sites, or two separates systems in equilibrium
(i.e. with equal values of $\mathcal{A}$). In this paper, we will
consider the first method, which then requires the proportionality
factor between the number of bulk and interface sites. Degeneracy
factors $g_{i}$ are given with respect to the total number of lattice
sites in the system, which are in fact divided into $N_{b}$ bulk
sites and $N_{\phi}$ interface sites: $N=N_{b}+N_{\phi}$. Assuming
that the interface is a spherical grain boundary, we are able to assign
specific values to $N_{b}$ and $N_{\phi}$ as a function of the grain
radius $r$:
\begin{equation}
\gamma=\dfrac{N_{\phi}}{N_{b}}=\frac{4\pi r^{2}V_{at}}{\frac{4}{3}\pi r^{3}S_{at}},\label{eq:LTE5}
\end{equation}
where $S_{at}$ and $V_{at}$ are the atomic surface and atomic volume,
respectively. In our case, Fe is body-centered cubic such that $V_{at}=a^{3}/2$
($a$ is the lattice parameter) and the GB plane that
we study has orientation $\left(111\right)$ hence $S_{at}=a^{2}\sqrt{3}$.
In the end, $\gamma=a\sqrt{3}/2r$. This quantity is useful to define $g_{i}$ coefficients for configurations specific to bulk or interface
environment because it is much easier to count segregation sites per interface site for instance. Thus, if we write $g_{i}^{\phi}$ and  $g_{i}^{b}$ the multiplicity per interface and bulk site, respectively, the $g_{i}$ coefficients that appear in Eq. \ref{eq:LTE4} are computed as:
\begin{equation}
    g_{i}=\dfrac{\gamma}{1+\gamma}g_{i}^{\phi}=\dfrac{1}{1+\gamma}g_{i}^{b}. \label{eq:gi2}
\end{equation}

Moreover, it is straightforward to generalize to more complex microstructure with multiple GB types, for instance: $N=N_{b}+N_{\phi_1}+N_{\phi_2}$ and we compute the $\gamma$ factor for each interface. Also, in Eq. \ref{eq:LTE5}, the $\gamma$ factor is defined as the surface over volume ratio of a sphere, but one could include a more sophisticated interface shape.

\subsection{Embrittlement potency\label{subsec:Embrittlement-potencies}}

Without any solute segregation, the energy of a system containing an interface reads:
\begin{equation}
E_{0}^{\phi}=NE_{m}+N_{\phi}\rho_{\phi}\gamma_{\phi}, \label{eq:E0}
\end{equation}
where $E_{m}$ is the energy of one matrix atom, $N$ is the total number of sites in the system, $\gamma_{\phi}$ is the interface energy per unit surface and $\rho_{\phi}$ is the interface area per interface atom. Note that $\rho_{gb}=\rho_{fs}=\rho$ and $N_{fs}=2N_{gb}$ because the fracture of a GB creates two free surfaces.
The ideal (reversible) work of separation per GB site ($W_{sep}^{0}$) quantifies the energy change as two grains are separated
from each other,
\begin{equation}
    W_{sep}^{0}=\dfrac{E_{0}^{fs}-E_{0}^{gb}}{N_{gb}}=\rho\left(2\gamma_{fs}-\gamma_{gb}\right). \label{eq:Wsep0}
\end{equation}

The embrittlement potency of a segregating element translates its ability to promote or reduce the probability of inter-granular fracture. It is commonly computed as the difference in segregation energy at the grain boundary and at the free surface, as originally proposed by Rice and Wang \citep{RiceWang1989, Wachowicz2011}: $EP=-\left(E_{gb}^{b}-E_{fs}^{b}\right)$ and the minus sign is added because in our convention, a positive binding energy represents an attractive configuration (see Eq. \ref{eq:Eb_defect}). When a single solute segregates to a single type of interface sites without interacting with other solutes, the solute embrittlement potency quantifies the difference in the ideal work of separation due to the presence of the solute. If the fracture occurs fast enough so that solutes do not have time to redistribute between bulk and free surfaces:
\begin{equation}
W_{sep}=W_{sep}^{0}-\left[X\right]EP, \label{eq:EP}
\end{equation}
where $\left[X\right]$ is the solute concentration per GB
site. Hence, a solute with a positive embrittlement potency (i.e. the solute is more stable at the free surface than at the GB) will promote
inter-granular failure as it will reduce the absolute value of $W_{sep}$.

\begin{figure}[ht]
\begin{centering}
\includegraphics[width=1\columnwidth]{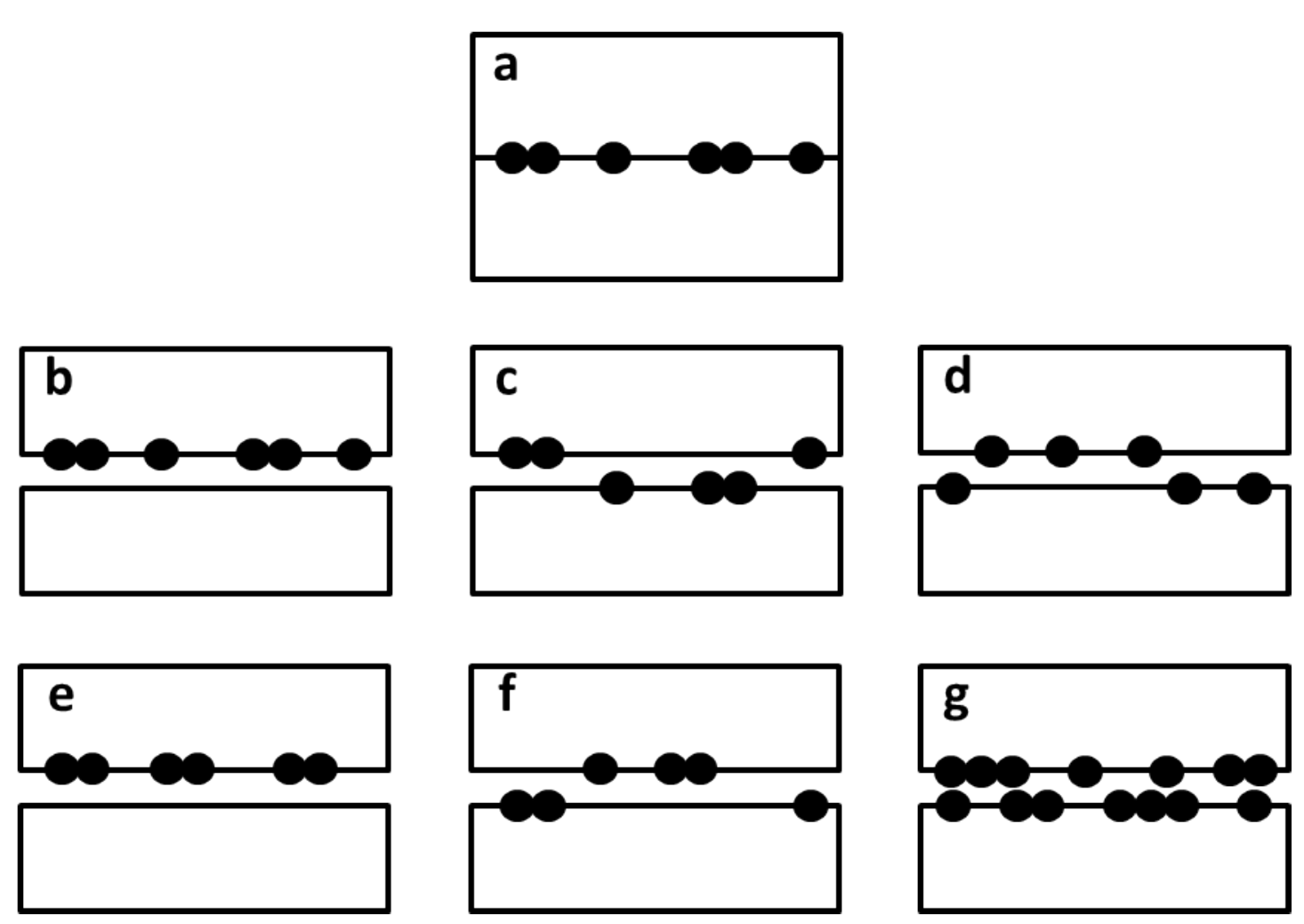}
\par\end{centering}
\centering{}\caption{\label{fig:fracture}Schematics of various scenarios of fracture
at a GB containing segregated solutes (a), into two free
surfaces (b-g). Various hypothesis are considered to quantify the
amount of solutes segregated to free surfaces (see text).}
\end{figure}

When various solutes interact and are located at various segregation sites, a generalized formula of the embrittlement potency is required. First of all, let us consider various fracture scenarios, as shown in Fig. \ref{fig:fracture}. Figure \ref{fig:fracture}a represents solutes segregated at the GB, at equilibrium where pairs and isolated solutes exist. Figures \ref{fig:fracture}b-\ref{fig:fracture}g represent different scenarios of GB separation into two surfaces. In Figs. \ref{fig:fracture}b-\ref{fig:fracture}d it is assumed that fracture is much faster than diffusion such that there is no solute equilibration at the surface. In Fig. \ref{fig:fracture}b, all solutes are located on the same surface, while they are evenly distributed between both surfaces in Fig. \ref{fig:fracture}c. If pairs and monomers do not interact (segregated solutes are sufficiently dilute), scenarios \ref{fig:fracture}b and \ref{fig:fracture}c will have the same energy, although scenario \ref{fig:fracture}c cannot result in flat surfaces. Figure \ref{fig:fracture}d shows a scenario similar to Fig. \ref{fig:fracture}c except that solute pairs are possibly separated during the fracture, which might lead to a different free surface energy. In Figs. \ref{fig:fracture}e-\ref{fig:fracture}g it is assumed that diffusion is fast compared with the fracture phenomenon, such that atomic species redistribute at free surfaces. In Fig. \ref{fig:fracture}e, all solutes are located on the same free surface, while they are evenly separated between both surfaces in Fig. \ref{fig:fracture}f. Both scenarios might not lead to the same energy because solute distribution depends on solute concentration. Contrary to previous cases, Fig. \ref{fig:fracture}g shows an example where the system is at fixed chemical potential, such that solutes are added or removed in order to equilibrate with the free surface. 

To evaluate the embrittlement potency, we will consider two fracture scenarios among those presented here (\ref{fig:fracture}b, \ref{fig:fracture}f). Indeed, as our model assumes a dilute solid solution with flat interfaces, scenarios \ref{fig:fracture}b and \ref{fig:fracture}c will give the same energy difference with respect to the GB configuration, and we have no way to quantify the distribution of solutes after fracture in Fig. \ref{fig:fracture}d, as the short-range order remaining at the free surfaces is unknown. Scenario \ref{fig:fracture}e could in principle be computed but it is cumbersome and it is not expected to provide much insight compared with scenario \ref{fig:fracture}f which is much easier to compute. Scenario \ref{fig:fracture}g requires a very slow fracture to enable long-range diffusion during the fracture event, which, in most cases, does not seem realistic. Moreover, because segregation energies to free surfaces are usually much higher than segregation energies to GBs (and this is the case in the Fe-P-H system, cf. Table \ref{tab:gbenergies}), such scenario would generate very large segregations to free surfaces, much larger than at GBs, thus a clear thermodynamic driving force for GB fracture, which translates into solute embrittlement.

To generalize the definition of the embrittlement potency, we use
the results from the LTE calculations to compute the work of separation. We consider the average energy of the interface at thermodynamic equilibrium as being representative of the system. The complete derivation of the following expressions are given in \ref{generalEP}. We define the generalized embrittlement potency ($GEP$) as the deviation from the ideal work of separation for pure materials, a deviation that is caused by a given concentration of one or several chemical species. For scenario \ref{fig:fracture}b, the concentration is conserved and fracture is fast with respect to the diffusion time scale of solutes, such that the free surface solute segregation is identical to the GB solute segregation:
\begin{equation}
    GEP_b=\sum_{i\in GB}\left(E_{i,fs}^{b}-E_{i,gb}^{b}\right)g_i^{gb}\exp\left(\beta\mathcal{A}_{i,gb}^{b}\right), \label{GEP_b}
\end{equation}
For scenario \ref{fig:fracture}f, the fracture is slow compared to diffusion time scales, allowing solutes to reach local equilibrium at the free surfaces, with the constraint of constant local concentration:
\begin{align}
    & GEP_f = \sum_{\alpha}\left[\alpha\right]\left(\mu_{\alpha}^{fs}-\mu_{\alpha}^{gb}\right) \nonumber \\
    & +\sum_{i\in GB} g_i^{\phi} \left(2E_{i,fs}^{b}\exp\left(\beta\mathcal{A}_{i,fs}^{b}\right)-E_{i,gb}^{b}\exp\left(\beta\mathcal{A}_{i,gb}^{b}\right)\right) \nonumber \\
    & +\sum_{i\in bulk} \dfrac{g_i^{b}E_{i,b}^{b}}{\gamma} \left((1-\gamma)\exp\left(\beta\mathcal{A}_{i,fs}^{b}\right)-\exp\left(\beta\mathcal{A}_{i,gb}^{b}\right)\right) , \label{GEP_f}
\end{align}
These expressions will be used in Sec. \ref{results} to compute the general embrittlement potency in each case.

\section{DFT computations \label{dft}}

\subsection{Bulk energies\label{subsec:Bulk-energies}}

First, we present our DFT results for bulk configurations. The purpose
of these calculations is to check the correct convergence of our DFT
calculations, and to quantify the interaction between P and H atoms
in the bulk, which may affect their distribution and thus their segregation
to GBs in fixed concentration samples. All calculations
were performed at fixed volume.

The vacancy formation energy in a $3\times3\times3$ cubic supercell
(54 atoms) of ferromagnetic body-centered cubic Fe was computed as:

\begin{equation}
E^{f}\left(vac\right)=E\left[53\mathrm{Fe}+vac\right]-\frac{53}{54}E\left[54\mathrm{Fe}\right],\label{eq:efv}
\end{equation}
and we found $E^{f}\left(vac\right)=2.19$ eV, in good agreement with
previous calculations (2.02-2.20 eV \citep{Messina2014}). To ensure
that our simulation cell was large enough we used the {\sc Aneto}
code which computes the elastic self-interaction of a point defect
from elastic constants and residual stress in the supercell \citep{VarvenneANETO}.
The energy of elastic interactions with supercell replicas amounts
to 0.017 eV for the vacancy, which is of the order of the error bar
resulting from our choices of DFT parameters. Hence, we consider that
this 54-atom supercell is large enough for bulk calculations.

The energy of a bulk system containing one substitutional P atom was
computed for 24- and 54-atom supercells. Both results agree within
0.02 eV and the energy due to elastic interaction with supercell replicas
amounts to 0.007 eV. We also computed the energy of an interstitial
octahedral P atom but the energy was 3.2 eV above that of substitutional
P atom (taking into account elastic corrections), slightly higher
than previous calculations (2.8 eV \citep{Fu2007} and 3.1 eV \citep{Domain2005}).
Therefore we will only consider substitutional P atoms hereafter. 

The H atom is more stable as an interstitial atom, and can occupy
both octahedral and tetrahedral positions. The tetrahedral site was
found more stable, 0.15 eV lower in energy than the octahedral site,
in agreement with previous calculations (0.13 eV \citep{JiangCarter}
and 0.12 eV \citep{Hayward2013}). Yet, as pointed out in Sec. \ref{subsec:Density-functional-theory},
it is important to account for zero-point energy (ZPE) effects for
light elements. This ZPE correction amounts to 0.14 eV and 0.24 eV
for the octahedral and tetrahedral sites, respectively, again in agreement
with previous results (0.12 eV and 0.24 eV, respectively\citep{Hayward2013}).
After ZPE corrections, the energy difference between tetrahedral and
octahedral sites reduces to 0.045 eV only, meaning that both sites must
be taken into account. The elastic interaction computed with {\sc Aneto}
amounts to 0.017 eV and 0.002 eV, respectively for the octahedral
and tetrahedral sites, again indicating that our 54-atom supercell
is large enough. The octahedral site is found unstable (second-order saddle point) and will not be included in our calculations. Therefore, the tetrahedral site is chosen as the reference H configuration to compute binding energies
(see Eqs. \ref{eq:Eb} and \ref{eq:Eb_defect}).

\begin{figure}[ht]
\begin{centering}
\includegraphics[width=1\columnwidth]{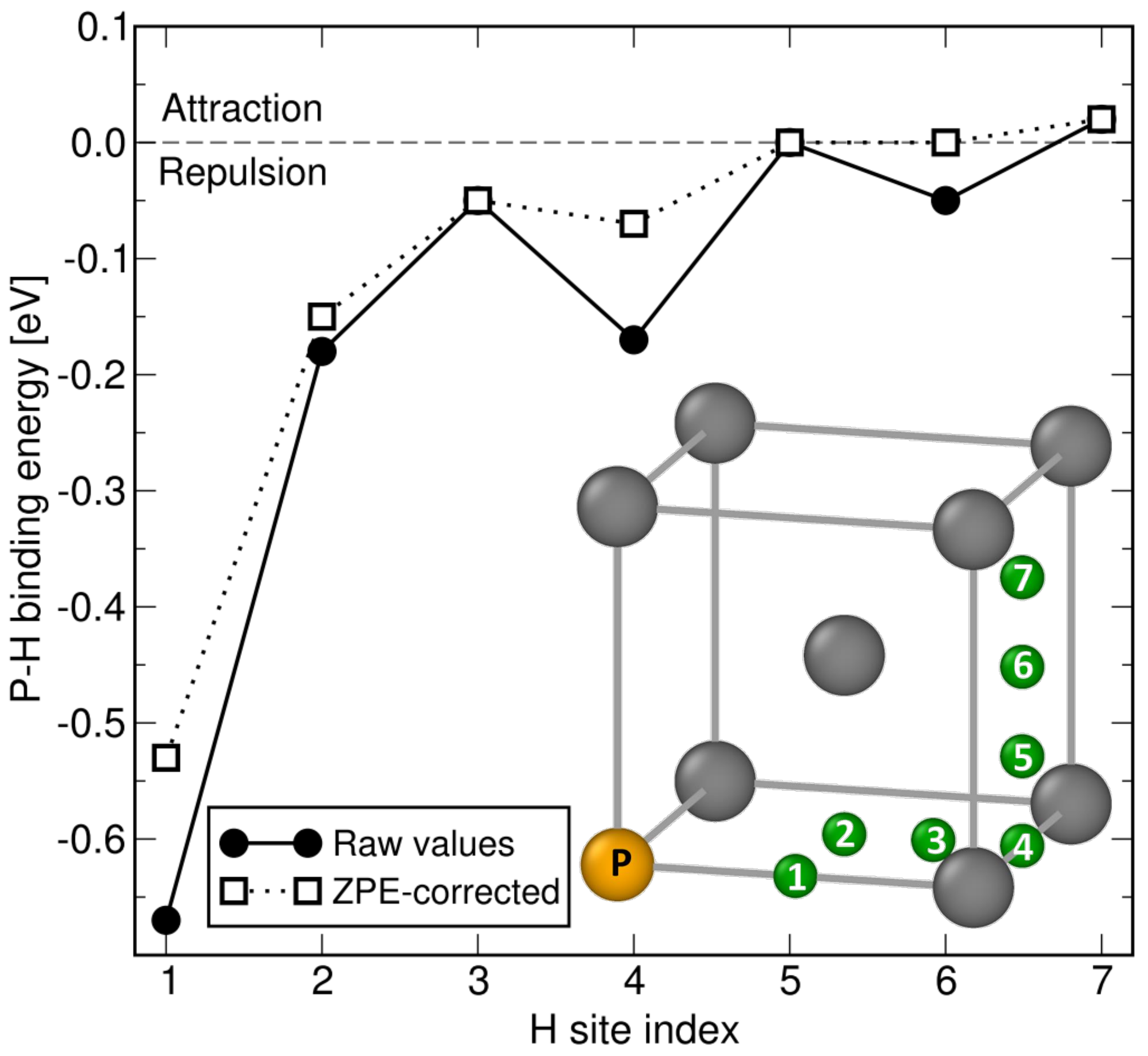}
\par\end{centering}
\caption{\label{fig:bulksites}Evolution of the P-H binding energy as a function
of the P-H configuration, labeled by the H site index. The various
P-H configurations considered in this study are pictured in the bottom-right
corner. The numbered green spheres show the position of the H atom
with respect to the P atom. Sites 1, 2, 4 and 6 are octahedral sites
and sites 3, 5 and 7 are tetrahedral sites.}
\end{figure}

Figure \ref{fig:bulksites} and Table \ref{tab:bulkenergies} show
the various P-H configurations and the associated binding energy.
At small distance, P and H have a negative binding energy, resulting
from a repulsive configuration. We also tested the tetrahedral site
located in between octahedral sites 1 and 2, but this configuration
spontaneously relaxes towards configuration 3. As the P-H distance
increases, binding energies converge towards 0, which is the expected
behavior with no long-range interactions. From this plot, we can estimate
the range of P--H interactions to be about one Fe lattice parameter,
with a strong repulsion at small P-H distances. The calculation of vibrational modes for H shows that configurations $\mathrm{PH_{1}}$ and $\mathrm{PH_{4}}$ correspond to second-order saddle-points, and configurations $\mathrm{PH_{2}}$ and $\mathrm{PH_{6}}$ correspond to first-order saddle-points. These configurations are therefore unstable and will not be included in the calculations.

\begin{table}[ht]
\begin{centering}
\begin{adjustbox}{width=\columnwidth}
\begin{tabular}{cccccc}
\hline 
Configuration & Site geometry & $g_{i}^{b}$ & $E^{b,u}$ & $E^{b}$ & Distance\tabularnewline
\hline 
\hline 
$\mathrm{PH}_{1}$ & Octahedral & 6 & -0.67 & -0.53 & 0.50a\tabularnewline
$\mathrm{PH}_{2}$ & Octahedral & 12 & -0.18 & -0.15 & 0.71a\tabularnewline
$\mathrm{PH}_{3}$ & Tetrahedral & 24 & -0.05 & -0.05 & 0.90a\tabularnewline
$\mathrm{PH}_{4}$ & Octahedral & 24 & -0.17 & -0.07 & 1.12a\tabularnewline
$\mathrm{PH}_{5}$ & Tetrahedral & 48 & -0.00 & -0.00 & 1.15a\tabularnewline
$\mathrm{PH}_{6}$ & Octahedral & 24 & -0.05 & -0.00 & 1.22a\tabularnewline
$\mathrm{PH}_{7}$ & Tetrahedral & 48 & +0.02 & +0.02 & 1.35a\tabularnewline
\hline 
\end{tabular}
\end{adjustbox}
\par\end{centering}
\caption{\label{tab:bulkenergies} Numerical values of binding energies for
P-H configurations depicted in Fig. \ref{fig:bulksites}.
The $g_{i}^{b}$ column corresponds to the number of symmetry equivalent
configurations per bulk lattice sites. $E^{b,u}$ is the uncorrected
binding energies, while $E^{b}$ is the ZPE-corrected binding energy
value. The last column shows the P-H distance in units of lattice parameter $a$.}
\end{table}

\subsection{Grain boundary and surface energies\label{subsec:Grain-boundary-energies}}

\begin{figure}[ht]
\begin{centering}
\includegraphics[width=1\columnwidth]{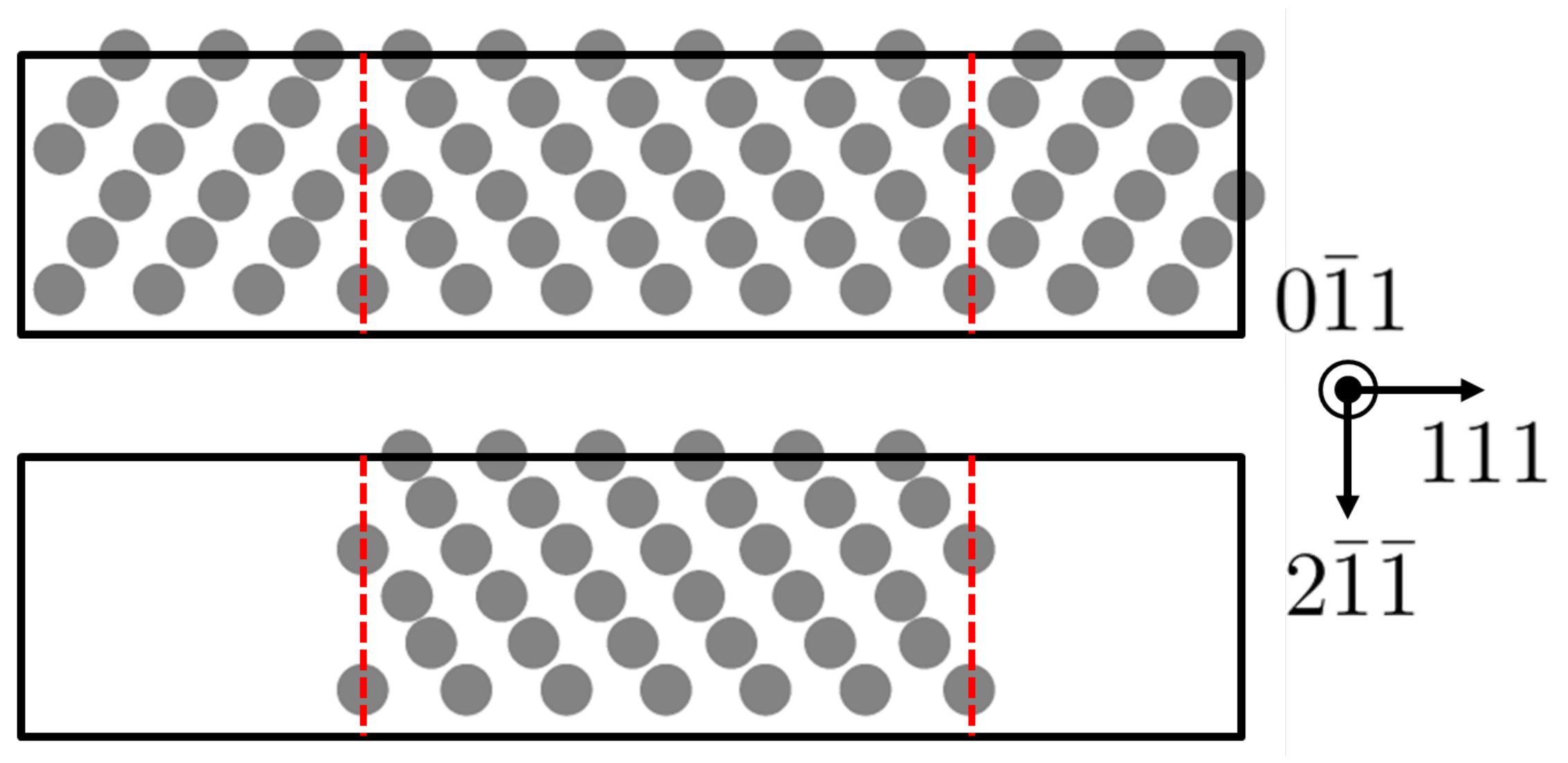}
\par\end{centering}
\caption{\label{fig:sideviews}Side views of the simulation cells used for
GB (top, 144 atoms) and free surface (bottom, 76 atoms)
calculations. The red dashed lines represent the GB or
free surface planes, perpendicular to the sheet plane.}
\end{figure}

\begin{figure}[ht]
\begin{centering}
\includegraphics[width=1\columnwidth]{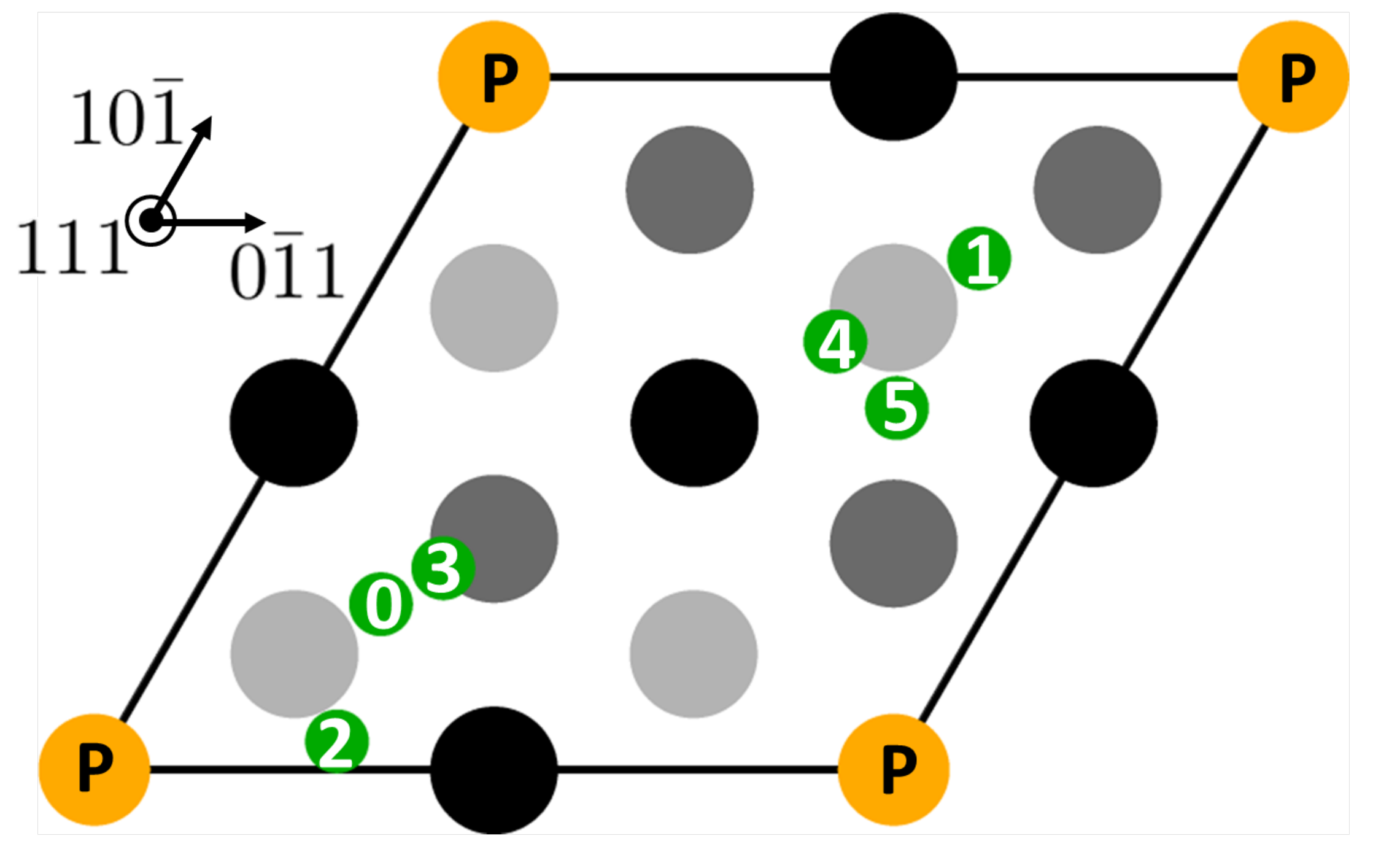}
\par\end{centering}
\caption{\label{fig:segsites}Top view of the $\left\{ 111\right\}$ interface
(GB plane or free surface). The green numbered circles
correspond to hydrogen positions, as in Table \ref{tab:gbenergies}.
Black circles are Fe atoms that are in the same plane as P atoms;
dark grey circles and light grey circles represent Fe atoms that are
respectively $a/2\sqrt{3}$ and $a/\sqrt{3}$ above and/or below the
plane, $a$ being the lattice parameter. To help visualization, atoms are replicated on the edges of the cells, but the supercell contain only one P atom in our calculations.}
\end{figure}

This section presents the P and H segregation energies to $\Sigma3$
$109.5\degree\left[0\bar{1}1\right]\left\{ 111\right\}$ GB and $\left\{ 111\right\} $ free surface, as well as P-H binding energies at these interfaces. The simulation cells are shown in Fig. \ref{fig:sideviews}. They contain 4 atoms per interface plane which is defined by two vectors of length 0.801 nm forming a 60\textdegree angle corresponding to $\left\langle110\right\rangle$-type directions (see Fig. \ref{fig:segsites}). 

The simulation cells were relaxed only in the direction perpendicular to the interface plane. If we had a very large box in the direction normal to the interface plane, and we would relax the simulation cell in all three directions, the relaxation would only occur in the direction normal to the interface plane, because a relaxation in a direction parallel to the interface plane would modify the inter-atomic distance for all bulk atoms and thus increase their energy, and as they would be much more numerous than interface atoms, such relaxation would not be energetically favorable. In our calculations we are limited in the size of the simulation cell we study. Even though we have checked that the simulation cell was large enough to avoid elastic interactions between both interfaces (10 mJ/m$^2$ variation in going from 12 to 24 atoms in between interfaces), the number of bulk atoms is not so large compared with the number of interface or near-interface atoms, such that some relaxation of the cell parallel to the interface plane might be energetically favorable. Hence, it seems more realistic to freeze the degrees of freedom parallel to the interface plane when relaxing the simulation cell. This may explain the difference in GB energy found with previous DFT calculations were the simulation cell was relaxed in all three directions (all values are given in J/m$^2$): 1.46 \citep{Mirzaev2016} and 1.52 \citep{Gao2009, Yamaguchi2010-1} whereas we find 1.23. This value is consistent with molecular static simulations using semi-empirical interatomic potentials, which are performed on much larger cells than DFT calculations: 1.25 \citep{Scheiber2016}, 1.31 \citep{Rajagopalan2013} and 1.55, 1.39, 1.30, 1.23, 1.22 using various potentials \citep{Gao2009}. We find the $\{ 111\}$ free surface energy at 2.07 (previous DFT calculations give 2.69 \cite{Yamaguchi2010-1}).

For the GB, we performed spin-polarized calculations with both collinear and non-collinear magnetic moments. Over all atoms, the maximum relative difference in magnetic moment was 0.6\% which is negligible. Hence we performed all subsequent calculations with collinear magnetic moments, which is cheaper in terms of computational cost. 

\begin{figure}[ht]
\begin{centering}
\includegraphics[width=0.7\columnwidth]{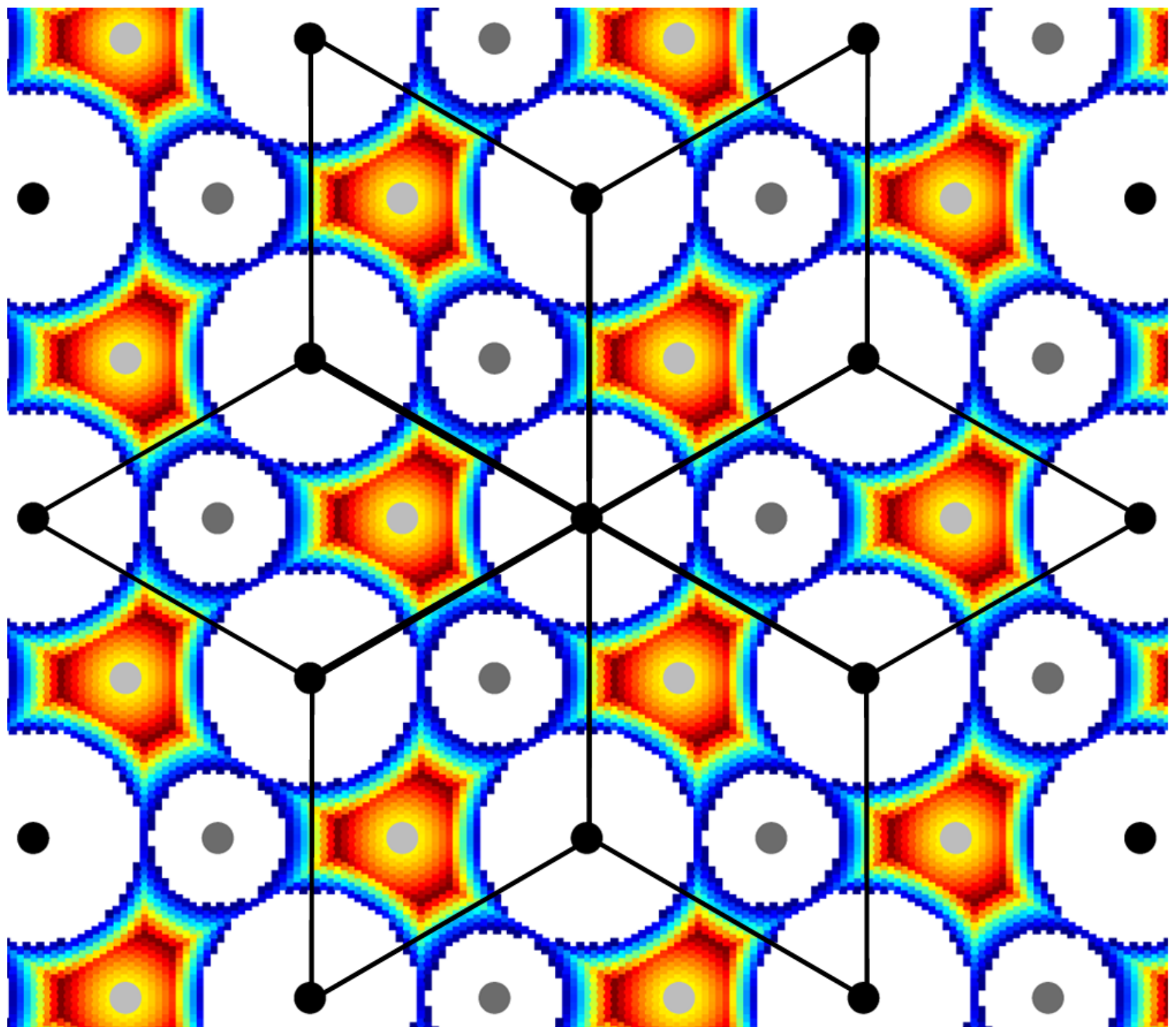}
\par\end{centering}
\caption{\label{fig:spaces}Top view of the $\left\{ 111\right\} $ interface
(GB plane of free surface). The black diamond-shaped lines
represent six unit supercells (each one being 1/4 of the actual supercell
that we used). For each point at the interface plane, the color shows the distance to the nearest Fe atom, dark red being 0.75a and dark blue being 0.56a, which is the Fe-tetrahedral site distance. Fe atoms are shown in black for interface atom, in grey and light grey for Fe atoms located $a/2\sqrt{3}$ and $a\sqrt{3}$ away from the interface plane, respectively.}
\end{figure}

To reduce the number of segregation sites to study we analyzed the symmetry of our simulation cell, and found an order 3 rotation symmetry around the $[111]$ axis as well as a symmetry plane normal to the $[110]$ direction. These symmetries are clearly seen in Fig. \ref{fig:spaces} where black-diamond shape lines show the unit supercell for this GB, and our actual supercell is composed of 4 unit supercells, as shown in Fig. \ref{fig:segsites}. Figure \ref{fig:spaces} additionally shows, for each point in the simulation cell, the distance to the closest Fe atom. This procedure aims at identifying potential interstitial segregation sites for H atoms. Red regions are the ones where there is the most space, so places where H atoms are expected to segregate. Hence we identified a collection of sites that are all equivalent when H is the only segregating species and infinitely dilute, represented by site 0 in Fig. \ref{fig:segsites}. We tried to insert P on this interstitial position but the calculation would not converge and the site was unstable. We conclude that P segregation is substitutional at this GB, in agreement with previous work \citep{Yuasa2011} but this might be true only for low temperatures \citep{Lejcek2016-P}. For P--H pairs, when P is placed at a substitutional site at the interface, interstitial sites in the GB plane become nonequivalent (numbered from 1 to 6, a H atom located at site 0 relaxes toward site 3 when P is added). Other sites were tested (red/yellow regions which denote available space in Fig. \ref{fig:spaces}) but relaxed towards one of these sites.

Table \ref{tab:gbenergies} shows the results of these DFT calculations in both GB and free surface for isolated P and H, P and H together but considered infinitely separated from each other, and P-H pairs with P being at a substitutional interface site and H being at an interstitial site numbered as in Fig. \ref{fig:segsites}. The last three columns show the difference between the P--H pair being at short distance and P and H being infinitely separated from each other, and thus quantifies the magnitude of the P--H interaction at the interface. First of all, the binding energy of H to the GB (including ZPE correction) is much higher than the binding energy of P to the GB. Hence, at similar bulk concentration levels, P is expected to be much less segregated to the GB than H. A similar comment holds for a free surface, even though the difference between the two species is much smaller. However, the embrittlement potency ($EP=E_{fs}^{b}-E_{gb}^b$) is positive for both species and much smaller for H than for P, which means that these solutes thermodynamically favor free surfaces and thus GB fracture, but P has a much stronger effect than H. We find an embrittlement potency of 0.25 eV/at for H--in good agreement with previous work: 0.26 eV/at \citep{Zhong2000, MATSUMOTO2011}, 0.33 eV/at \citep{Yamaguchi2010} and 0.41 eV/at \citep{Mirzaev2016}--and 1.48 eV/at for P while previous DFT results show widely spread values: 1.63 eV/at \citep{Yuasa2013}, 0.79 eV/at \citep{Wu1994}, 0.17 eV/at \citep{Geng2001}.

Looking at co-segregation effects in detail, all P--H configurations are repulsive at the GB and only one configuration is attractive at the free surface. Thus, P--H interactions will tend to lower slightly the segregation levels at the GB and increase slightly the segregation levels at free surfaces, meaning that they provide an additional thermodynamic driving force for GB fracture. Looking at the last column of Table \ref{tab:gbenergies}, the embrittlement potency difference is positive for 3 out of the 5 configurations tested, confirming the additional embrittlement of the GB caused by the short-range order between P and H. Note that configuration $\mathrm{PH}_{4}$ is in fact found unstable (first order saddle point) when computing vibrational modes for the H atom, and this configuration will not be included in our calculations.

\begin{table}[ht]
\begin{centering}
\begin{adjustbox}{width=\columnwidth}
\begin{tabular}{cccccccccc}
\hline 
 & $g_{i}^{\phi}$ & $E_{gb}^{b,u}$ & $E_{gb}^{b}$ & $E_{fs}^{b,u}$ & $E_{fs}^{b}$ & $EP$ & $\Delta E_{gb}^{b}$ & $\Delta E_{fs}^{b}$ & $\Delta EP$\tabularnewline
\hline 
\hline 
P & 1 & 0.16 & 0.16 & 1.64 & 1.64 & 1.48 & - & - & - \tabularnewline
$\mathrm{H}_{0}$ & 3 & 0.52 & 0.64 & 0.82 & 0.95 & 0.31 & - & - & - \tabularnewline
P+$\mathrm{H}_{0}$ & - & 0.68 & 0.80 & 2.46 & 2.59 & 1.79 & +0.00 & +0.00 & +0.00 \tabularnewline
$\mathrm{PH}_{1}$ & 3 & 0.66 & 0.77 & 2.44 & 2.56 & 1.79 & -0.02 & -0.03 & +0.00 \tabularnewline
$\mathrm{PH}_{2}$ & 6 & 0.47 & 0.56 & 2.23 & 2.33 & 1.77 & -0.23 & -0.26 & -0.02 \tabularnewline
$\mathrm{PH}_{3}$ & 3 & 0.67 & 0.79 & 2.55 & 2.68 & 1.89 & -0.01 & +0.09 & +0.10 \tabularnewline
$\mathrm{PH}_{4}$ & 3 & 0.35 & 0.48 & 2.22 & 2.34 & 1.86 & -0.32 & -0.25 & +0.07 \tabularnewline
$\mathrm{PH}_{5}$ & 6 & 0.65 & 0.76 & 2.44 & 2.57 & 1.81 & -0.04 & -0.02 & +0.02 \tabularnewline
\hline 
\end{tabular}
\end{adjustbox}
\par\end{centering}
\caption{\label{tab:gbenergies}P--H binding energies at the GB and free surface.
The first two lines are binding energies of a single atom at each interface, which are related to segregation energies. The third line is the sum of the first two. The last 5 lines are binding energies between P and H at each interface. $g_{i}^{\phi}$ is the degeneracy per interface site--accounting for configurational entropy--and the $u$ superscript indicates that the ZPE correction is not taken into account (stands for "uncorrected") while for ZPE-corrected binding energies, Eq. \ref{eq:zpe} was used. $EP$ is the embrittlement potency for this configuration only, and the last three columns are computed with respect to the third line, to emphasize co-segregation effects.}
\end{table}

 In short, at equivalent bulk concentrations, P segregation to the GB is lower than that of H, but at equivalent segregation levels, P makes the GB more brittle than H. When P and H are included simultaneously in the system, segregation levels are expected to be mostly as if there was no interaction between P and H but some configurations cause more GB embrittlement than isolated P and H. The next step consists in computing how much of these P--H pairs form as a function of temperature and nominal solute concentrations and thus how the embrittlement potency of the system is affected. To this end, we will use the LTE model presented in Sec. \ref{methods}.

\section{Effect of co-segregation on embrittlement\label{results}}

\subsection{Effect of grain size on segregation\label{subsec:grainsize}}

One of the nice features of the LTE model is that it is straightforward to take into account the effect of grain size, which translates into a contribution to configurational entropy: the larger the grain, the lower the number of interface sites at constant total number of sites, hence the lower the configurational entropy associated with interface sites. In this study, the ratio between the number of interface sites and the total number of sites is computed in the ideal case of spherical grains (cf. Eq. \ref{eq:LTE5}).

\begin{figure}[ht]
\begin{centering}
\includegraphics[width=1\columnwidth]{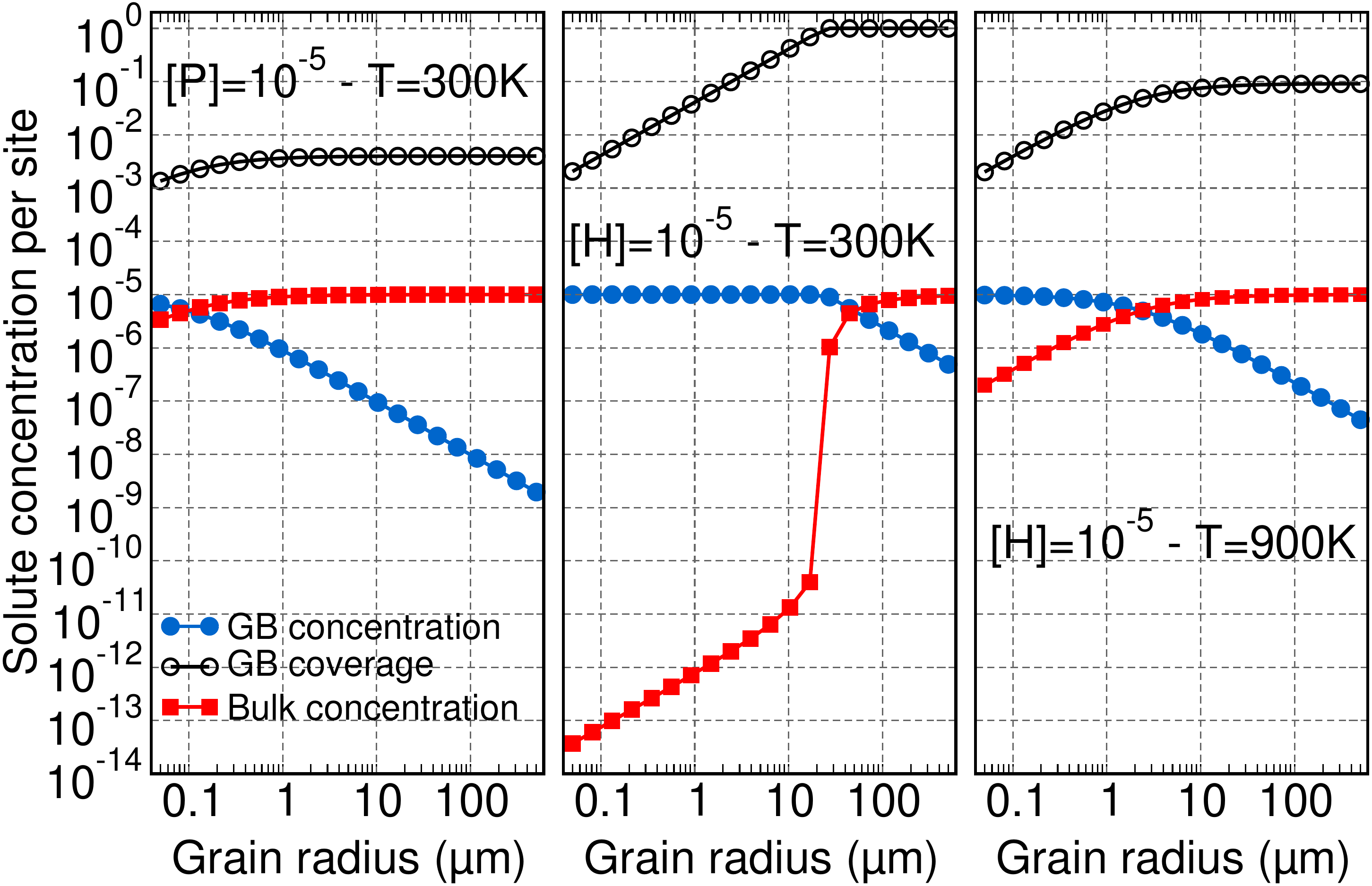}
\par\end{centering}
\caption{Solute concentrations as a function of grain radius for various temperatures and nominal concentrations. Three concentrations are shown: GB concentration (blue circles) which is the number of solutes at GBs divided by the total number of sites (bulk+GB); bulk concentration (red squares) which is the number of solutes in the bulk divided by the total number of sites; GB coverage (open black circles) which is the number of solutes at GBs divided by the number of GB sites. Note that P--H interactions are not taken into account in these calculations.
\label{fig:Reffect}}
\end{figure}

Figure \ref{fig:Reffect} shows the effect of grain radius for isolated species (no P--H pair formation is allowed). As shown in Table \ref{tab:gbenergies}, the H binding energy to the GB is about 4 times larger than the P binding energy to the GB. Hence H segregation is expected to be much more important than P segregation. This is indeed found from the two first sub-figures, where solute concentrations in the GB plane and in the bulk were computed at T=300 K for a 10 appm nominal solute concentration. At identical temperature and nominal solute concentrations, GB segregation is higher for H than for P, and for both species it decreases with increasing grain radius. This behavior is also expected because higher grain size favor bulk sites over GB sites in terms of configurational entropy. 

A less-intuitive conclusion drawn from these calculations is that GB coverage increases and then seems to saturate with increasing grain size. GB coverage is the number of solutes at GBs per GB sites, while GB concentration is the number of solutes at GBs per total number of sites in the system. As grain radius increases, the average solute energy in the system--its chemical potential--increases because more and more solutes are added to the bulk where they are less stable than at GBs. At some point, most solutes are in the bulk such that increasing the grain radius further does not really affect the average solute energy in the system, hence the saturation of GB coverage. Figure \ref{fig:Reffect} shows that GB coverage and bulk concentration have the same variations, which comes from the fact that they are both proportional to the exponential of the solute chemical potential. This variation of GB coverage with grain size is clearly seen from  LTE expressions, considering only one species (H) without geometrical frustration effects, to keep the discussion simple. Also we assume that $1/(1+\gamma)\simeq1$, which is true except for very small grains (a few nm). Defining $S_{H}=3\exp\left(F_{gb}^{b}\left(H^{gb}\right)/k_{B}T\right)$ and
$Y=\exp\left(\mu_{\mathrm{H}}/k_{B}T\right)$, the total H concentration is the sum of bulk H atoms and H atoms segregated at the GB:
\begin{equation}
\left[H\right]=\dfrac{1}{1+\gamma}6Y+\dfrac{\gamma}{1+\gamma}YS_{H}\simeq Y\left(6+\gamma S_{H}\right).
\end{equation}
The GB coverage by H atoms is expressed as:
\begin{equation}
\theta_{H}=YS_{H}=\dfrac{\left[H\right]S_{H}}{6+\gamma S_{H}}=\dfrac{\left[H\right]}{\dfrac{6}{S_{H}(T)}+\dfrac{\sqrt 3}{2r}}.\label{eq:thetaH0}
\end{equation}
Hence at given H nominal composition and temperature, higher grain radius $r$ results in high GB coverage until a saturation value is reached, $\theta _H\simeq [H]S_H/6$. 

The second sub-figure in Fig. \ref{fig:Reffect} (H segregation at T=300K) is particularly interesting because GB coverage saturates at 1, which means that all interstitial sites in the GB plane are occupied by H atoms. When concentrations reach high levels on a specific type of sites, one must be cautious with LTEs, because geometrical frustration effects must be added. They consist in adding excitation states where solutes are non-interacting. These states have a negligible probability at low solute concentrations but become necessary when concentration rise above $\simeq1\%$. Generally speaking it is a difficult task to compute these frustration effects. Yet, we can perform the exact calculation when solutes are considered non-interacting, as it is done in  \ref{sec:Appendix} and \ref{generalEP}. In this Section, all calculations were performed taking into account geometrical frustration in this ideal case for solutes at interfaces, treated in a mean-field way. This way of dealing with interface concentration issues is not exact but allows to keep calculations not too complicated. Note that only geometrical frustrations are treated in a mean-field way, segregation being treated exactly and independently for each type of segregation site. \ref{app_counter} presents the derivation of the mean-field geometrical frustration and an example of the equations solved in this study.

\subsection{Effect of temperature on co-segregation}
\begin{figure}[ht]
\begin{centering}
\includegraphics[width=1\columnwidth]{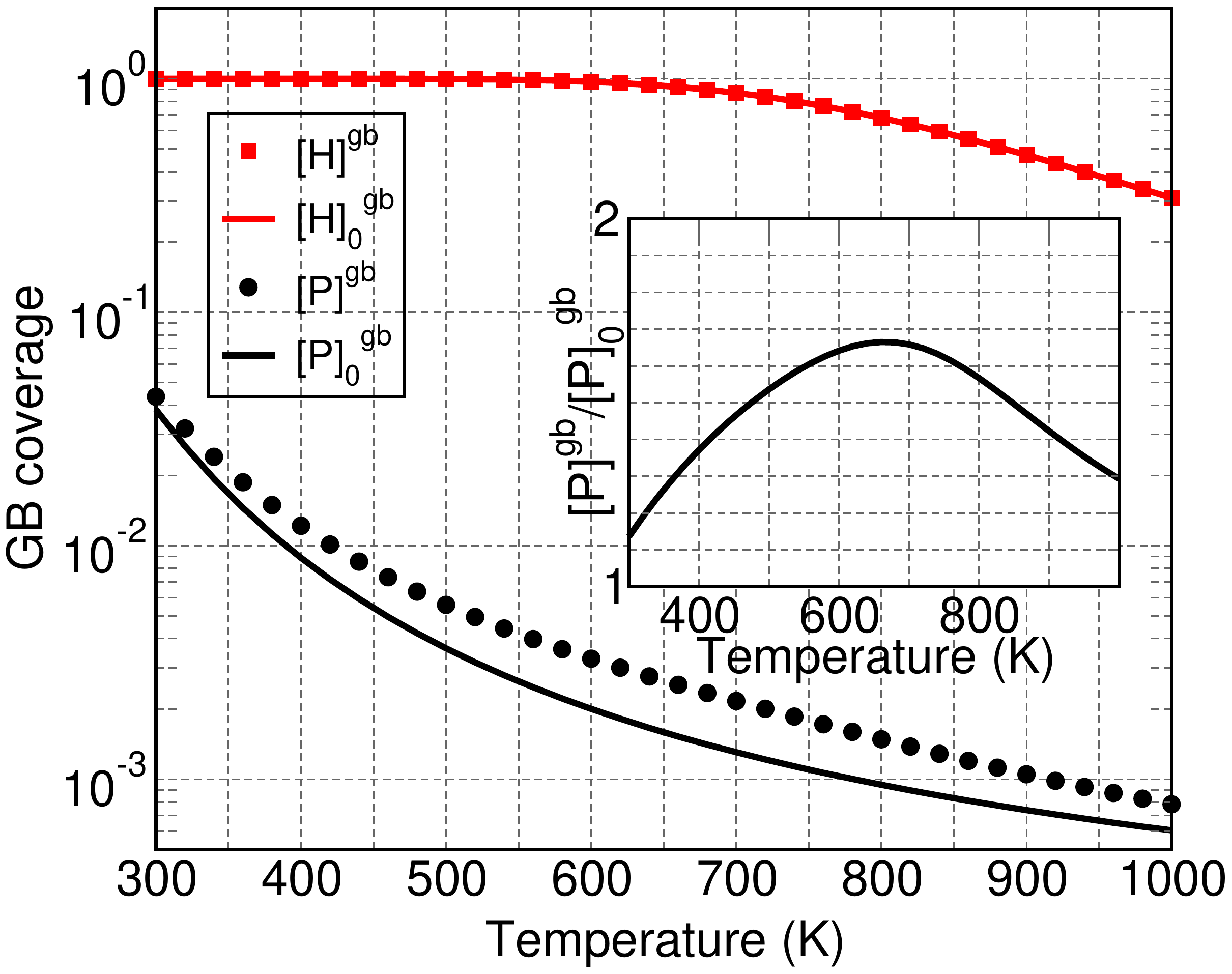}
\par\end{centering}
\caption{GB coverage as a function of temperature for H solutes (red) and P solutes (black) when they are non-interacting (solid lines, denoted by the 0 subscript in the legend) or interacting via the formation of P--H pairs (symbols). Grain radius is fixed at $10\mu$m and nominal concentrations are set to 100 appm for both species. The inset shows the ratio of P GB coverage, with and without H in the system.\label{fig:temp}}
\end{figure}

Figure \ref{fig:temp} shows how GB coverage evolves as a function of temperature in two cases: either P and H solutes are treated separately, i.e. they do not interact (solid lines); or P--H interactions--summarized in Tables \ref{tab:bulkenergies} and \ref{tab:gbenergies}--are accounted for (symbols). Hence, the difference between symbols and solid lines represents the effect of co-segregation.

The decrease of GB coverage with temperature is trivial as segregation is proportional to the exponential of the segregation energy over $k_{B}T$. It is interesting to note the asymmetry in H and P behavior: P--H interactions increase P segregation at GBs by a factor lower than 2, and this effect  is temperature dependent, with a maximum at $T\simeq 650K$ (see inset). On the other hand, H segregation is not affected by 100 appm of P, whatever the temperature, and additional calculations not shown here confirm that H segregation is unaffected by the presence of P in the system. This can be understood from Fig. \ref{fig:Reffect}: at a grain radius of $10\mu$m, H concentration at GBs is 1-2 orders of magnitude higher than that of P. Hence, most P atoms can pair with H atoms while leaving most H atoms isolated, i.e. without changing the H chemical potential. The reverse asymetric behavior is expected at free surfaces because P segregation energy is much higher than H segregation energy. 

Binding energies being all repulsive for P--H pairs at GBs (Table \ref{tab:gbenergies}), the increase of P segregation in the presence of H is due to P--H pairs that are stabilized by configurational entropy only. The inset shows that the co-segregation effect--quantified by the ratio of P segregated concentration with and without H in the system--presents a maximum at T$\simeq$650 K. As temperature increases, the configurational entropy contribution to the free energy of the system becomes more and more significant, and H concentration at GBs is more or less constant, close to a full coverage of the GB plane. At some temperature, the H coverage at GBs decreases, which reduces the probability of forming P--H pairs, and therefore the stabilization of P atoms by configurational entropy also decreases.

\begin{figure}[ht]
\begin{centering}
\includegraphics[width=1\columnwidth]{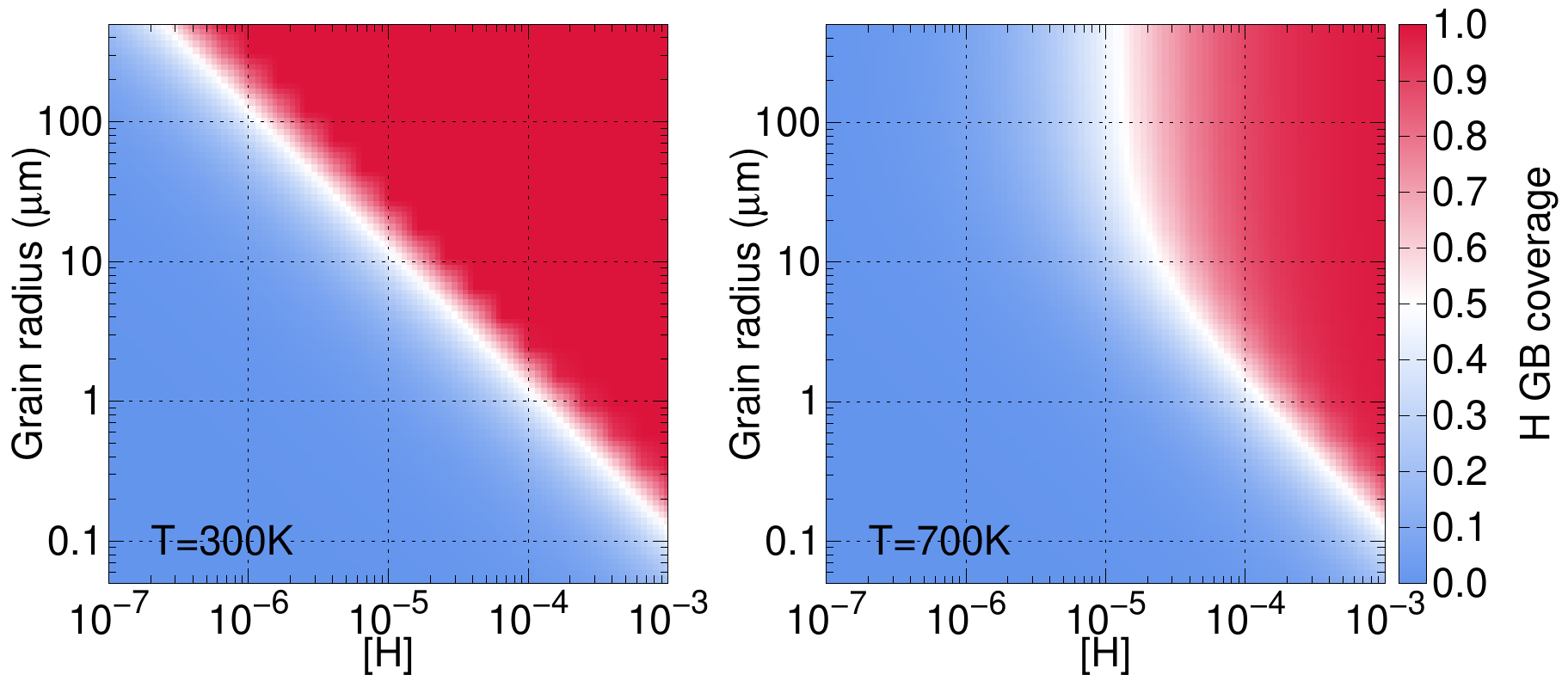}
\par\end{centering}
\caption{H GB coverage as a function of H nominal concentration and grain radius for two temperatures: T=400K and T=700K. P nominal concentration is set to 50 appm for these plots.\label{fig:hsat}}
\end{figure}

To summarize, co-segregation effects on segregation are mainly independent of P concentration and triggered by a large H concentration at GBs, which increases P segregation. It is therefore interesting to take a look at H segregation at GBs which depends on H nominal concentration, temperature and grain radius. Figure \ref{fig:hsat} shows the H GB coverage (H concentration per GB site at the interface plane) as a function of grain radius and H nominal concentration for two temperatures. 

At low temperature, constant GB coverage contours appear as straight lines in Fig. \ref{fig:hsat}, either vertical or with a slope equal to -1. This may be explained from the LTE expressions. Taking the natural logarithm of Eq. \ref{eq:thetaH0} we get:
\begin{equation}
\ln\left[H\right]=\ln\left(\dfrac{6\theta_{H}}{S_{H}}\right)+\ln\left(1+\dfrac{\sqrt{3}S_H}{12r}\right), \label{eq:thetaH}
\end{equation}
from which we identify two limiting behaviors: if $\sqrt{3}S_{H}\ll12r$ (large grain radius and/or high temperature), Eq. \ref{eq:thetaH} simplifies to  $\ln\left[H\right]\simeq\ln\left(6\theta_{H}/S_{H}\right)$, which does not depend on the grain radius and should indeed give vertical constant coverage contour in Fig. \ref{fig:hsat}; on the contrary, if $\sqrt{3}S_{H}\gg12r$ (low grain radius and/or low temperature) we get a linear relation with a slope equal to -1 between the logarithms of grain radius and nominal H concentration $\ln\left[H\right]\simeq\ln\left(\sqrt{3}\theta_{H}/2\right)-\ln\left(r\right)$. Hence, Eq. \ref{eq:thetaH} qualitatively explains the trends observed in Fig. \ref{fig:hsat} and shows that the comparison between the H segregation energy over $k_BT$ and the grain radius is the key quantity to predict H coverage and thus the possible P--H co-segregation effects.

\subsection{Combined effects of grain radius, temperature and H nominal concentration on co-segregation}

\begin{figure}[!ht]
\begin{centering}
\includegraphics[width=1\columnwidth]{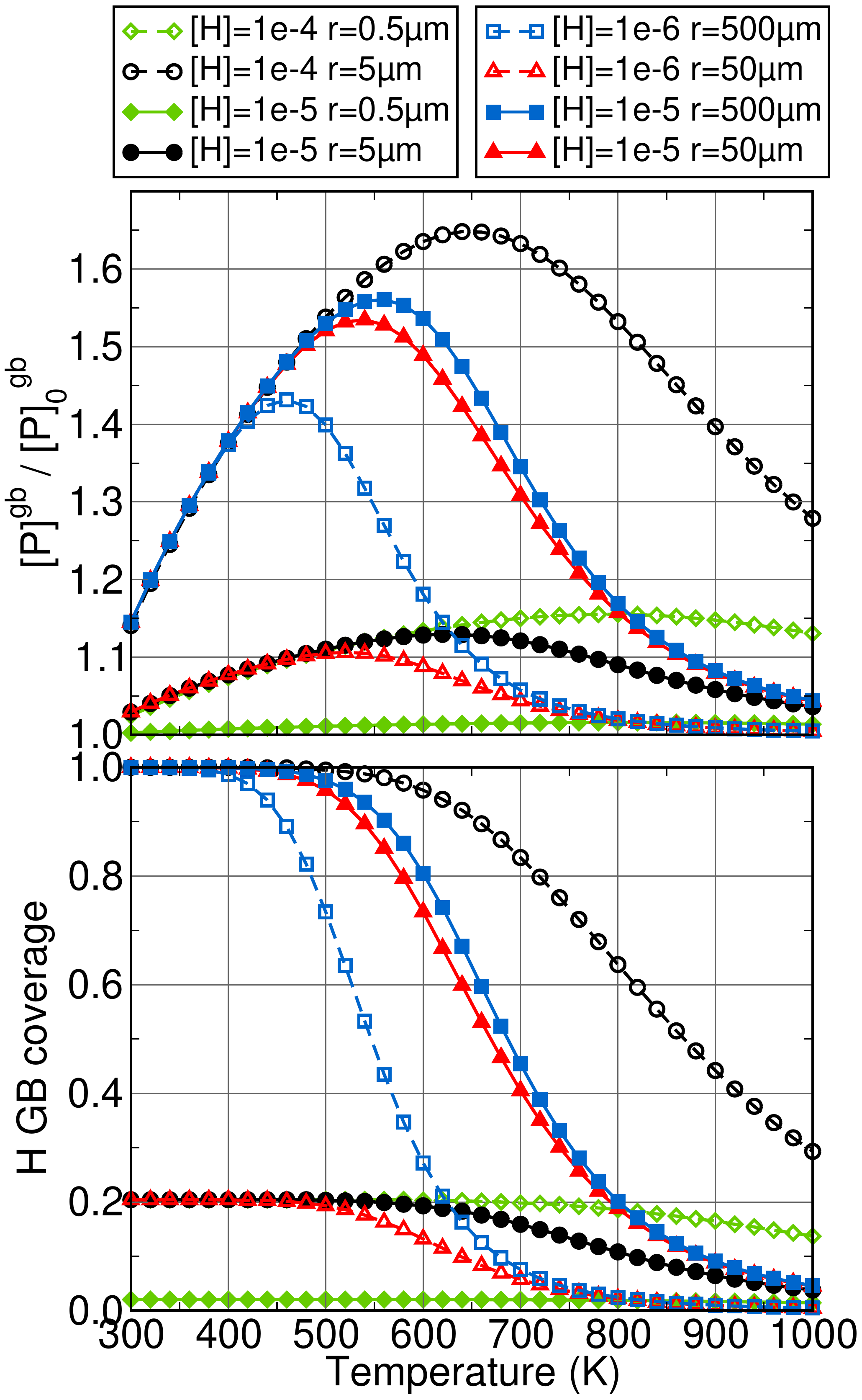}
\par\end{centering}
\caption{(Top) $[P]^{gb}/[P]^{gb}_{0}$ ratio as a function of temperature, which measures how H increases the segregation of P atoms to GBs; (Bottom) GB coverage of H atoms as a function of temperature. These plots were obtained at a constant P nominal concentration of 50 appm, for various grain radii and H nominal concentration. Filled and empty symbols are obtained for a one order of magnitude difference in H concentration, while going from green, to black, to blue, to red curves, grain radius increases by one order of magnitude each time.\label{fig:cosegp}}
\end{figure}

As in the inset of Fig. \ref{fig:temp}, the magnitude of co-segregation effects is expressed as the ratio between the computed P concentration at the GB and the P concentration obtained when H is removed from the calculation (hence no P--H interactions, denoted by subscript 0). The additional segregation of P to GB due to H is a function of temperature, grain radius and H nominal concentration. In this section we want to understand the interplay between these parameters. To this end, Fig. \ref{fig:cosegp} presents the $[P]^{gb}/[P]^{gb}_{0}$ ratio as a function of temperature for various combinations of grain radius and H concentration. The lower plot shows the corresponding GB coverage of H, which is a function of the three same parameters. The curves that we obtain have the same shape as the one in Fig. \ref{fig:temp}, with a maximum located more or less at the point where H concentration at the GB becomes lower than unity.

Focusing first on the lower plot of Fig. \ref{fig:cosegp}, the curves show three values of H coverage at room temperature. This behavior is explained in Eq. \ref{eq:thetaH} by the comparison between $S_H$ and $r$. By comparing some of the curves, we see that increasing the grain radius by one order of magnitude while decreasing the nominal concentration by the same amount does not change the low temperature coverage, but modifies the temperature evolution of $\theta _H$.

Looking now at the top plot of Fig. \ref{fig:cosegp}, the curves also seem to form three groups of curves, the ones with a maximum $[P]^{gb}/[P]^{gb}_{0}$ ratio between 1.4 and 1.7, the ones with a maximum ration between 1.1 and 1.2, and finally the one with almost constant ratio, and these "groups" correspond to the ones identified for H coverage. Generally speaking, increasing the grain radius increases the co-segregation effects (higher maximum of the curves) and translates the maximum ratio towards lower temperatures; increasing the H nominal concentration increases the co-segregation effect and translates the maximum ratio towards higher temperatures. Hence, these two variables do not have the same effect, because they do not affect the temperature evolution of the H coverage in the same way. The comparison between the top and bottom plots of Fig. \ref{fig:cosegp} clearly demonstrates that the magnitude of co-segregation effects is directly linked with the fraction of interface sites occupied by H atoms. 

\subsection{Generalized Embrittlement Potency: rapid fracture at constant concentration}

\begin{figure}[ht]
\begin{centering}
\includegraphics[width=1\columnwidth]{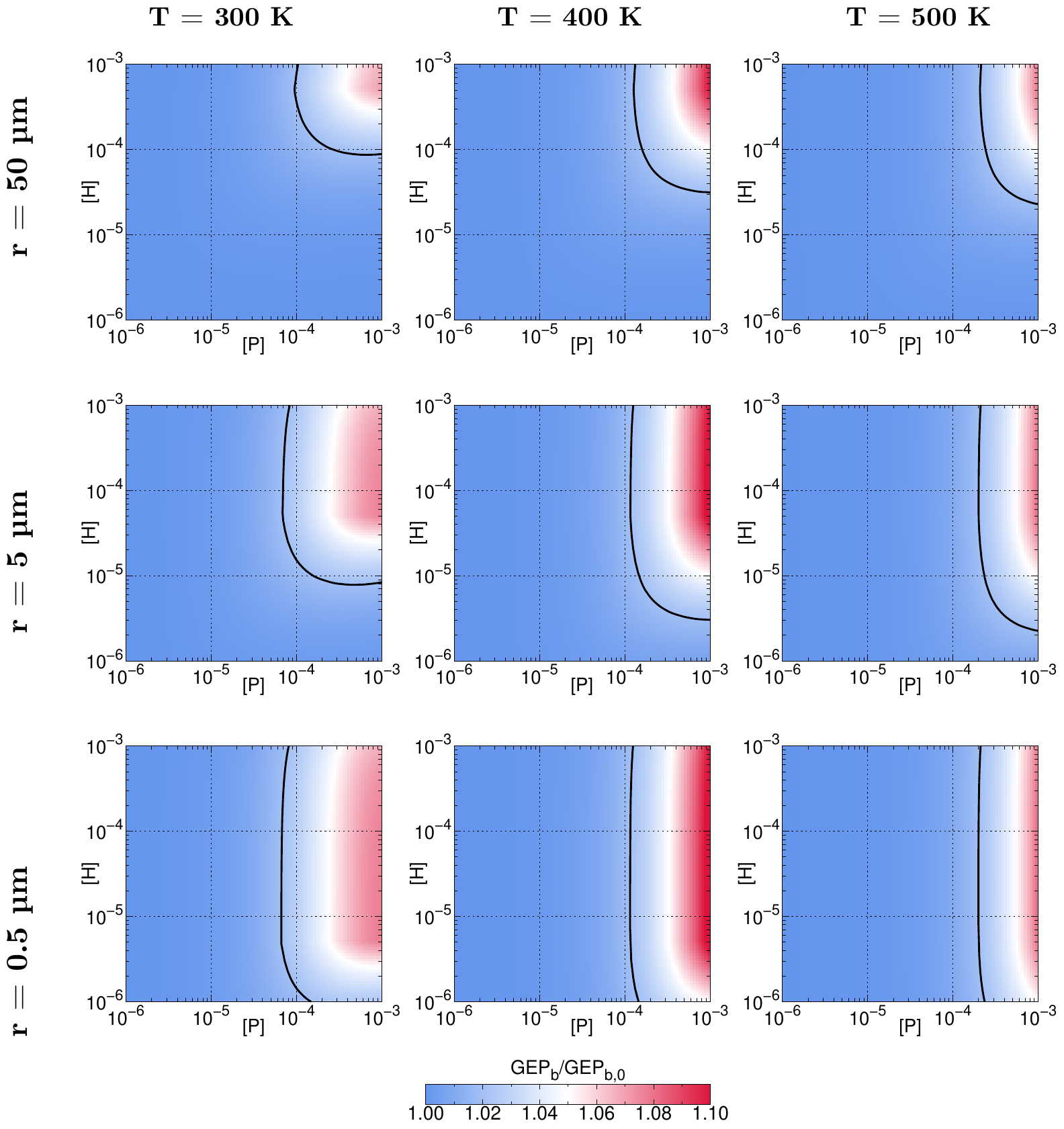} 
\par\end{centering}
\caption{Generalized embrittlement potency for scenario \ref{fig:fracture}b (cf. Eq. \ref{GEP_b}). These plots show the ratio between the GEP obtained in the FePH alloy normalized by the GEP that would be obtained if P and H were not interacting. Hence it directly quantifies the effect of co-segregation on the embrittlement potency. The black line shows the contour at a ratio of 1.02.\label{fig:GEPb}}
\end{figure}

We now discuss the general embrittlement potency (GEP) obtained for scenario \ref{fig:fracture}b where segregated solutes are assumed to be immobile during the fracture phenomena, such that they are simply transferred from the GB to the free surface, and the GEP is expressed by Eq. \ref{GEP_b}. Figure \ref{fig:GEPb} shows the ratio between the GEP and the one that would be obtained without P--H interactions, so as to characterize co-segregation effects on embrittlement potency. This ratio is shown as a function of solute nominal concentrations for various temperatures and grain sizes. From previous paragraphs, we know that P--H interactions lead to larger concentrations of segregated P for H coverage close to unity, hence low temperature and/or large grain radius, such that the qualitative trends shown in Fig. \ref{fig:GEPb} are expected. In addition to larger quantities of segregated P, \textit{ab initio} calculations showed that P--H pairs mostly lead to increased embrittlement potency (cf. Table \ref{tab:gbenergies}), adding up with increased segregation levels and leading to larger GEP, i.e. more brittle interfaces. Nevertheless, note that the GEP potency increase due to co-segregation remains quite low, always lower than 10\% and occurs only at fairly high nominal concentrations of P (at least 100 appm). This is at variance from the effect on segregation levels which was mainly independent from the concentration of segregated phosphorus. In the P--H system, the embrittlement potency is essentially an average of the embrittlement potency of each species, weighted by their concentration. Hence, if H is much more concentrated than P at the interface, the contribution of P atoms on the embrittlement potency will be essentially negligible. If H concentration at the interface is low, then there is no co-segregation effect. 

\subsection{Generalized Embrittlement Potency: slow fracture at constant concentration}

\begin{figure}[ht]
\begin{centering}
\includegraphics[width=1\columnwidth]{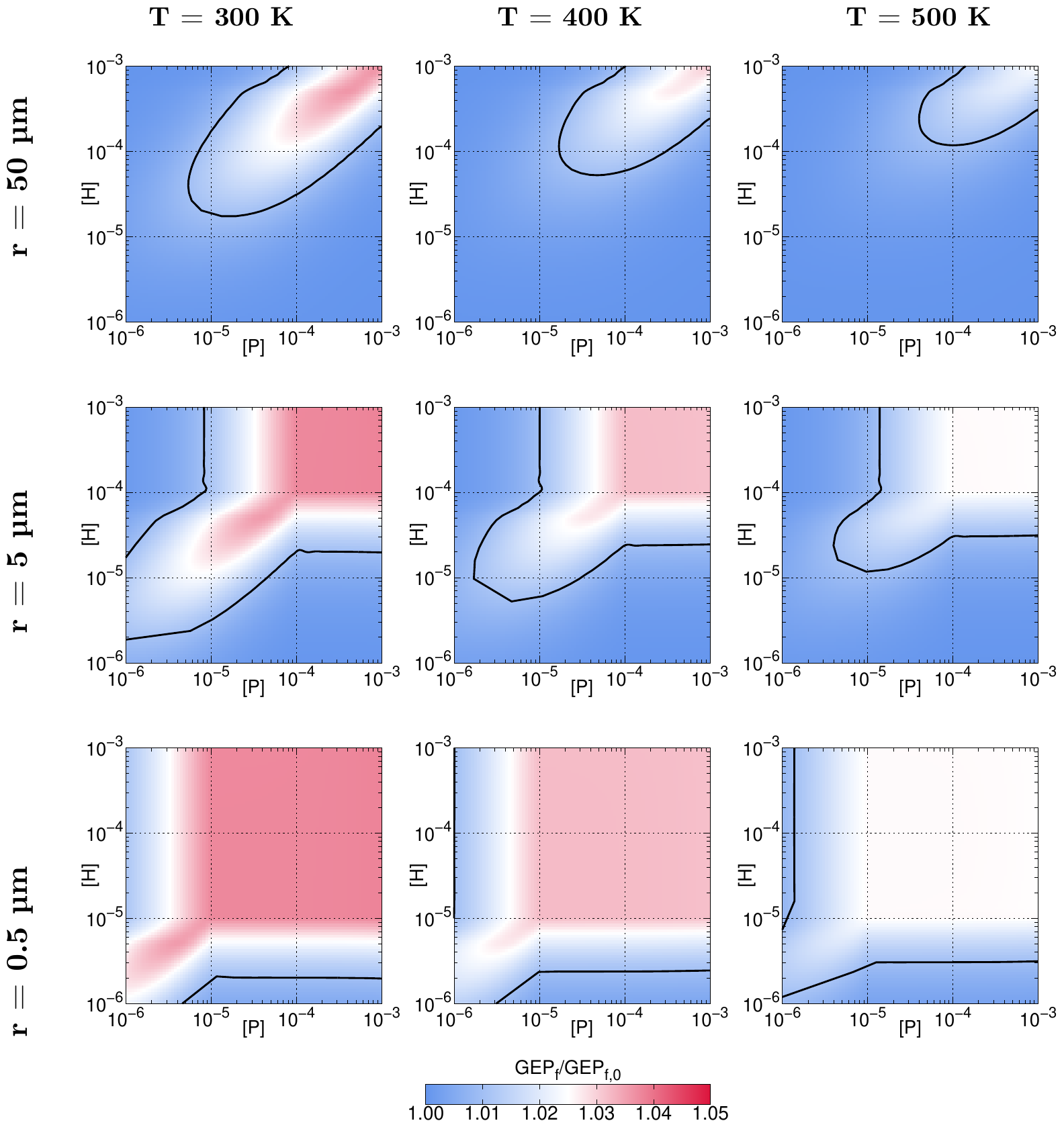}
\par\end{centering}
\caption{Generalized embrittlement potency for scenario \ref{fig:fracture}f (cf. Eq. \ref{GEP_f}). These plots show the ratio between the GEP obtained in the FePH alloy normalized by the GEP that would be obtained if P and H were not interacting. Hence it directly quantifies the effect of co-segregation on the embrittlement potency. The black line shows the contour at a ratio of 1.01.\label{fig:GEPf}}
\end{figure}

The general embrittlement potency obtained for scenario \ref{fig:fracture}f characterizes what happens during a slow fracture, when segregated solutes may diffuse locally and equilibrate at the free surface during the fracture. The corresponding GEP is expressed by Eq. \ref{GEP_f}. As shown in Fig. \ref{fig:GEPf}, the effect of co-segregation on embrittlement is lower in this scenario than in the one describe in the previous paragraph, as in the range of parameters investigated, the GEP do not exceed a 5\% increase due to co-segregation, even though it occurs at lower nominal concentrations. The plots are also more symmetric than the ones obtained for fracture scenario \ref{fig:fracture}b, because, as H would increase P segregation at GBs, P increases H segregation at free surfaces, and this feature is now accounted for since solutes are equilibrated at free surfaces. Hence for each species there is some threshold concentration value above which co-segregation translates into increased GEP, either because the GB plane is covered with H atoms, or because the free surface plane is covered with P atoms. When both species are simultaneously in concentrations close to these threshold values, the increased GEP is also observed.

\section{Discussion about current model limitations \label{discussion}}

The aim of this paper was mainly to introduce the new LTE statistical model for solute segregation at interfaces. The application to P--H co-segregation effects in $\alpha$-Fe was limited by the amount of computational time available for DFT calculations.

The first way to improve the model is to perform additional DFT calculations, either on other sites around the GB because some out-of-plane sites may be found very stable \citep{Yuasa2011, Wang2016}. Also, a similar type of work should be performed on other GBs because a real material contains various types of GBs \citep{Morito2006}. Moreover, highly coherent GB such as first-order twins considered in this study may exhibit huge property changes if the GB plane is misaligned by only a few degrees \citep{RAABE2014}.
These types of complex structures--for instance non-symmetric GBs--are now becoming accessible, especially using molecular dynamics approaches based on semi-empirical interatomic potentials \citep{Scheiber2016, Frolov2013, Huber2017}. Note that having a statistical model such as the one proposed in this paper that is able to treat each segregation site individually but also collections of segregation sites (i.e. short-range order) is particularly useful for grain boundaries in ordered structures, e.g. oxides \citep{Lejcek2018-OpenQuestion}.

Another point which is crucial--especially for comparison with experiments--is to take into account temperature effects. For instance, DFT calculations at 0 K indicate substitutional segregation of phosphorus while there is experimental evidence for interstitial segregation at higher temperature \citep{Lejcek2016-P}. The discrepancy may be due to the temperature dependence of the segregation energy (entropy contribution) and magnetism evolution as the Curie temperature is approached. More generally speaking about finite temperature effects, segregation entropy is important to consider for relevant modeling/experiment comparison\citep{Lejcek2014, Lejcek2018-OpenQuestion}. In our LTE model, we can easily use a free energy for each configuration. As a matter of fact, we already included configurational entropy and a first-order approximation of vibrational entropy for H atoms in the present work. Other forms of entropy (e.g. electronic, magnetic and vibrational) can sometimes be computed with state-of-the-art DFT tools, but these remain computationally expensive, especially for complex GB structures. Temperature also affects the lattice parameter, magnetic ordering and GB structure which are all non trivial contributions. Because DFT calculations cannot consider all these contributions altogether, a closely bound experimental/numerical work is probably the best way to obtain meaningful data  on solute segregation at GBs.

One limitation of LTE models is their inability to treat phase transitions in the case of solute precipitation at the GB. There is a workaround though, but two separate LTE calculations must be performed--one with a reference free interface and the other with a reference precipitate phase at the GB--and then both systems must be put in equilibrium to solve concentrations and/or chemical potentials\citep{Barouh2014}. The main difficulty in performing LTE calculations is the computation of the so-called "counter-terms", i.e. the geometrical multiplicity of states that are dissociated\cite{Clouet2007}. These states are usually negligible for dilute concentrations (typically $<1at\%$ for a given type of site). For more concentrated cases these terms should be added, but it is always possible to check whether or not the next term in the expansion is negligible. Hence, one can safely use LTE expressions in more and more concentrated states without knowing the full (infinite) partition function of the system.

Finally, we were only focused on equilibrium segregation, while real materials contain multiple segregating elements, each having their own kinetic properties \citep{Morvan2017}. Thus out-of-equilibrium segregations may arise, for instance during quenching \citep{anthony:1970, anthony:1975}, under irradiation \citep{Nastar2012, Ardell2016} or because of GB migration \citep{Cahn1962}. Note that in all these cases the elastic field around the GB modifies the diffusion properties of solutes \citep{Vattre2016} and thus the segregation.

This rather long list of model limitations does not mean that DFT calculations are useless to study GB segregation, but rather points towards some critical points to keep in mind when using such models and comparing them with experiments.

\section{Conclusion}

The main contribution of this paper is the development of the low-temperature expansion model to study interface segregation. The strength of this model is its general formulation which makes it suitable to address various segregation problems, for instance taking into account multi-species and multi-sites per species segregation, as well the effect of grain size on segregation.

To demonstrate the applicability of the LTE formulation, we applied it to the study of P--H co-segregation effects in $\alpha$-Fe twin boundaries. We performed \textit{ab initio} calculations to obtain the input data necessary for the LTE model, mainly binding and segregation energies. 

We showed the importance of grain size on GB coverage, especially for small grains. The classic McLean segregation isotherm \citep{McLean1957} is basically restricted to the large grain size limit, where the solute chemical potential is given by the bulk solution energy. P and H solutes show asymmetric co-segregation behavior at the GB plane. H segregation levels are insensitive to the presence of P, while H atoms can slightly increase P segregation to GBs due to configurational entropy. For this to happen, the GB plane must be almost saturated with H atoms, to maximize the probability of forming P--H pairs. Hence, the H coverage at GBs is the key quantity, and depends on the grain radius, temperature and H nominal concentration. The situation is reversed at free surfaces where H may be stabilized by P atoms. 

A general formula for embrittlement potency is proposed, and we show that co-segregation can lead to increased embrittlement potency, up to 10\% depending on solute concentrations, temperature, grain size and the fracture scenario considered (comparison between fracture and local diffusion time scale).

Larger co-segregation effects would be obtained if P and H showed attractive interaction in the GB plane. This might be the case in other GBs, or for other solutes at a twin boundary. 

Because H easily saturate GBs, future work on segregation at $\alpha$-Fe twin boundaries should focus on concentration effects, introducing other clusters (H--H interactions for instance) and out-of-plane configurations. For other types of GBs, the strain field around the GB may favor these out-of-plane configurations. Finally, let us stress that the ideal work of interfacial separation is predicted to have an important but probably not exclusive role in controlling embrittlement \citep{RiceWang1989}: for instance, McMahon emphasizes the dynamic effect of H as an explanation to GB embrittlement \citep{McMahon2004}. Hence, equilibrium segregation studies provide necessary but not sufficient insight into embrittlement phenomena.

\section*{Acknowledgments}
The authors gratefully acknowledge the various computing centers where calculations were performed for this study: the HPCaVe center at UPMC-Universit\'e, the CENTAURE cluster at \'Ecole Nationale Sup\'erieure des Mines de Saint-\'Etienne, the P2CHPD center at Universit\'e de Lyon and the CALMIP center at Universit\'e de Toulouse  (UMS CNRS 3667). Part of this study was funded by the project GIBBS-ANR-15-CE30-0016, and the authors acknowledge financial support from the NEEDS-SEDI project and D. Conn\'etable for his contribution in getting access to computing resources.

\appendix
\section{Discussion about the reliability of DFT calculations for solutes with low solubility limits\label{sec:Appendix0}}

\textit{Ab initio} methods have their drawbacks, mostly related to the complexity of properly taking into account temperature effects, and the limited number of atoms that can be included in a simulation box. Regarding the latter, let us comment on some question that was raised about the reliability of \textit{ab initio} calculations for solutes with small solubility limits \citep{Lejcek2013, Lejcek2018-OpenQuestion}. We do not question the discrepancy between experimentally measured and computed segregation energy that was identified by the authors, but the "low solubility limit" argument is not convincing, mainly because it is not an intrinsic parameter of a solute in a matrix: the solubility limit depends on the phases in the material, not only on the thermodynamically stable ones but on the ones that are actually formed. So even though a solute with a solubility around 1 appm is severely supersaturated in a 100-atom simulation cell, it cannot precipitate because at best it will form a lattice with more than 1 nm between each atom, which is usually not the stable phase required to impose this solubility limit. The convergence of the energy of the system with the cell size demonstrates that there is no interaction between a solute and its replicas such that in the simulation cell, the solute cannot be out-of-equilibrium because solubility limit is not really defined. In our opinion, the discrepancy is rather due to complicated temperature effects and experimental difficulties in measuring segregation energies, all of which being emphasized for the solutes with "small solubility limits". But the \textit{ab initio} calculation itself is correct at $T=0$ K.

\section{Equivalence between LTE and McLean formalisms\label{sec:Appendix}}

In this Appendix, we demonstrate the equivalence between the low-temperature expansion formalism that is used in this paper, and the standard McLean model which is commonly used to compute the amount of segregated species \citep{Langmuir1918,McLean1957,Hondros1977,Lejcek2014}. In the Langmuir-McLean model, there is only one type of solute, segregating at one type of site (single segregation energy) and segregated atoms do not interact with one another. Hence, the solute-GB binding energy of state $i$ containing $n_{i}^{\alpha}=i$ solutes $\alpha$ is $E_{i}^{b}=iE^b_{\alpha}$, where $E^b_{\alpha}$ is the segregation energy of species $\alpha$ at the GB. 

From Eq. \ref{eq:LTE2}, we need to compute the degeneracy for each state to be able to compute the partition function of the system. Fortunately, for the specific case we are interested in, it is possible to give a general formula for each $G_i$ factor and write the infinite series of the partition function. The change in $G_i$ values as a function of the number of solutes in the system is mainly related to exclusion effects, because two solutes cannot occupy the same site. These exclusion effects are taken into account via "counter-terms" in the LTE formalism, which corresponds to $G_i$ factors for non-interacting of dissociated configurations. Also note that all solutes are not distinguishable hence the division by the number of permutations for $i$ solutes.

\begin{equation}
G_{1}=N,\label{eq:-1}
\end{equation}

\begin{equation}
G_{2}=\dfrac{N\left(N-1\right)}{2!},\label{eq:-2}
\end{equation}

\begin{equation}
G_{3}=\dfrac{N\left(N-1\right)\left(N-2\right)}{3!},\label{eq:-3}
\end{equation}

\begin{equation}
G_{i}=\dfrac{N\left(N-1\right)\cdots\left(N-i+1\right)}{i!},\label{eq:-4}
\end{equation}
The total energy of each state is given by $i\left(E_{i}^{b}+\mu_{\alpha}\right)$. Hence, defining $X=\exp\left(\left(E^{b}_{\alpha}+\mu_{\alpha}-\mu_{m}\right)/k_{B}T\right)$, Eq. \ref{eq:LTE2} becomes:
\begin{align}
\mathcal{A} & =\mathcal{A}_{0}-k_{B}T\ln\left(1+\sum_{i\neq0}\dfrac{N\left(N-1\right)\cdots\left(N-i+1\right)}{i!}X^{i}\right)\nonumber \\
& = \mathcal{A}_{0}-k_{B}TN\sum_{i\neq0}\dfrac{(-1)^{i-1}}{i}X^{i} \nonumber \\
& =\mathcal{A}_{0}-k_{B}TN\ln\left(1+X\right),\label{eq:demo1}
\end{align}
and the second equality makes use of the linked-cluster theorem \citep{Ducastelle1991}.

Matrix $m$ and solute $\alpha$ concentrations at the interface $\phi$
are obtained from the derivatives of Eq. \ref{eq:demo1} with respect
to the corresponding chemical potentials:

\begin{align}
\left[m\right]^{\phi} & =1-\frac{X}{1+X}\nonumber \\
\left[\alpha\right]^{\phi} & =\frac{X}{1+X},\label{eq:demo2}
\end{align}
such that $\left[m\right]^{\phi}+\left[\alpha\right]^{\phi}=1$ and
: 
\begin{equation}
\dfrac{\left[\alpha\right]^{\phi}}{\left[m\right]^{\phi}}=\exp\left(\frac{E^{b}_{\alpha}+\mu_{\alpha}-\mu_{m}}{k_{B}T}\right).\label{eq:demo3}
\end{equation}
The exact same work can be performed in the bulk rather than at the
interface, the only difference being that there is no segregation
energy in the bulk: $E^{b}_{\alpha}=0$. Thus, we recover the Langmuir-McLean segregation isotherm:

\[
\dfrac{\left[\alpha\right]^{\phi}}{1-\left[\alpha\right]^{\phi}}=\dfrac{\left[\alpha\right]^{b}}{1-\left[\alpha\right]^{b}}\exp\left(\frac{E^{b}_{\alpha}}{k_{B}T}\right).
\]

\section{Generalized embrittlement potency \label{generalEP}}
\subsection{Average energy of a system}
In this Appendix we derive the expressions presented in Sec. \ref{subsec:Embrittlement-potencies}. To generalize the definition of the embrittlement potency, we use the results from the LTE calculations to compute the work of separation. We consider the average energy of the interface at thermodynamic equilibrium as being representative of the system.
Let $\beta=1/k_{B}T$ and $\mathcal{A}_{i}=E_{i}-\sum_{\alpha}n_{i}^{\alpha}\mu_{\alpha}$. Then the partition function of the system reads $Z=\sum_iG_i\exp(-\beta\mathcal{A}_i)$, which is nothing but the sum of the probability of each micro-state. The average energy of the system (including chemical potential contributions in the grand-canonical ensemble) is:
\begin{align}
\left\langle \mathcal{A}\right\rangle & =\dfrac{\sum_{i}\mathcal{A}_{i}P_{i}}{\sum_{i}P_{i}}=\dfrac{\sum_{i}\mathcal{A}_{i}G_{i}\exp\left(-\beta\mathcal{A}_{i}\right)}{Z}=-\dfrac{1}{Z}\dfrac{\partial Z}{\partial\beta} \nonumber \\ & =-\dfrac{\partial\ln Z}{\partial\beta}=\dfrac{\partial\left(\beta\mathcal{A}\right)}{\partial\beta}=\mathcal{A}+\beta\dfrac{\partial\mathcal{A}}{\partial\beta}.\label{eq:-5}
\end{align}

Let us define the binding grand potential of a micro-state, defined with respect to reference state $i=0$: $\mathcal{A}_{i}^{b}=-\left(\mathcal{A}_{i}-\mathcal{A}_{0}\right)=E_{i}^{b}+\sum_{\alpha}\delta n_{i}^{\alpha}\mu_{\alpha}$.
Using the expression from Eq. \ref{eq:LTE3} and these notations:
\begin{equation}
    \mathcal{A}=\mathcal{A}_{0}-\dfrac{N}{\beta}\sum_{i\neq0}g_i\exp\left(\beta\mathcal{A}^b_i\right), \label{eq:reLTE}
\end{equation}
leading to:
\begin{equation}
\left\langle \mathcal{A}\right\rangle =\mathcal{A}_{0}-N\sum_{i\neq0}\mathcal{A}_{i}^{b}g_{i}\exp\left(\beta\mathcal{A}_{i}^{b}\right),\label{eq:-8}
\end{equation}
which is conveniently rewritten as:
\begin{equation}
\dfrac{\left\langle \mathcal{A}\right\rangle }{N}=\dfrac{E_{0}}{N}-\sum_{i\neq0}E_{i}^{b}g_{i}\exp\left(\beta\mathcal{A}_{i}^{b}\right)-\sum_{\alpha}\mu_{\alpha}\left[\alpha\right],\label{eq:-10}
\end{equation}
where $\left[\alpha\right]$ is the total concentration of species $\alpha$ in the system, given by Eq. \ref{eq:LTE4}. The transition from Eq. \ref{eq:-8} to Eq. \ref{eq:-10} is exact as long as all entropy contributions in $\mathcal{A}_i^b$ are temperature-independent, as the configurational entropy for instance. In our calculations, we consider temperature-dependent vibrational entropy for H atoms (cf. Eq. \ref{eq:Fvib}) but for simplicity we still use Eq. \ref{eq:-10} to compute the general embrittlement potency.

\subsection{Comparison with the ideal embrittlement potency in the dilute case}
With this expression of the average energy of the system, let us recover the basic expression of the embrittlement potency in the case where there is only one segregating species, one segregation site, and the interface concentration of solute is very low.
\begin{equation}
\left\langle \dfrac{\mathcal{A}}{N}\right\rangle =\dfrac{E_{0}}{N}-E^{b}\dfrac{N_{\phi}}{N}\exp\left(\beta\left(E^{b}+\mu_{\alpha}\right)\right)-\mu_{\alpha}\left[\bar{\alpha}\right],\label{eq:-11}
\end{equation}
If all the atoms segregated at the GB stay at free surfaces without
short-range order re-organization (fast fracture, as in scenario \ref{fig:fracture}b or \ref{fig:fracture}c),
then the probability of each configuration\textendash roughly $\exp\left(\beta\left(E_{i}^{b}+\mu_{\alpha}\right)\right)$
is identical at the GB and at the free surface. Also the chemical
potential is not really defined once the system is fractured because the system is not at equilibrium. Yet, if we assume that there are more solute atoms in the bulk than in the GBs or free surfaces-\textendash because grains are large enough so that the number of bulk sites is much larger than the number of interface sites\textemdash then we can assume that the solute chemical potential is mainly given by bulk solute atoms such that $\mu_{\alpha}^{gb}\simeq\mu_{\alpha}^{fs}$. For scenario \ref{fig:fracture}b, we will always make this assumption since the chemical potential is ill-defined after fracture. The ideal work of separation per GB site is:

\begin{equation}
W_{sep}=\left\langle \dfrac{\mathcal{A}}{N_{gb}}\right\rangle ^{fs}-\left\langle \dfrac{\mathcal{A}}{N_{gb}}\right\rangle ^{gb}=W_{sep}^{0}-X_{gb}EP_{\alpha},\label{eq:-14}
\end{equation}
with $X_{gb}=\exp\left(\beta\left(E_{gb}^{b}+\mu_{\alpha}\right)\right)$ is the concentration of solute atoms at the interface, sometimes called the interface coverage (this expression is valid for low coverage only) and the embrittlement potency for solute $\alpha$ is defined as in Sec. \ref{subsec:Embrittlement-potencies}: $EP_{\alpha}=E_{fs}^{b}-E_{gb}^{b}$. If the solute is more stable at the GB than at the free surface ($E_{gb}^{b}>E_{fs}^{b}$), the embrittlement potency will be negative and solute segregation increases $W_{sep}$ meaning it strengthens the GB. 

\subsection{Comparison with the ideal embrittlement potency including geometrical frustration effects}
Now let us do a comprehensive calculation for this specific case of one type of solute segregating at one type of interface site, taking full account of concentration effect. The expressions for $G_{i}$
coefficients are the ones shown in \ref{sec:Appendix} (Eq. \ref{eq:-4}), with $N=N_{gb}$ and we want
to compute the ideal work of separation per unit GB site, using
$g_{i}=\lim_{N_{gb}\rightarrow0}\left[G_{i}/N_{gb}\right]$. The energy
of state $i$ where $n_i^{\alpha}=i$ solute atoms have been added is $E_{i}^{b}=iE^{b}$ because solutes are non-interacting. 
\begin{equation}
\dfrac{\left\langle \mathcal{A}\right\rangle }{N_{gb}}=\dfrac{E_{0}}{N_{gb}}-E^{b}\sum_{i\neq0}\left(-1\right)^{i+1}\dfrac{i\left(i-1\right)!}{i!}X^{i}-\sum_{\alpha}\mu_{\alpha}\left[\bar{\alpha}\right].\label{eq:-15}
\end{equation}
Knowing that
\begin{equation}
\sum_{i\neq0}\left(-1\right)^{i+1}\dfrac{i\left(i-1\right)!}{i!}X^{i}=\frac{X}{1+X},\label{eq:-16}
\end{equation}
we find the following expression, which is similar to Eq. \ref{eq:-14} for low $X_{gb}$ values:
\begin{equation}
W_{sep}=W_{sep}^{0}-\frac{X_{gb}}{1+X_{gb}}EP_{\alpha}.\label{eq:-17}
\end{equation}
\subsection{Generalized embrittlement potency expressions}
We now derive the expressions presented in Sec. \ref{subsec:Embrittlement-potencies} to define the generalized embrittlement potency for scenarios b, f and g in Fig. \ref{fig:fracture}. We define the generalized embrittlement potency ($GEP$) as the deviation from the ideal work of separation for pure materials, a deviation that is caused by a given concentration of one or several chemical species:
\begin{equation}
 GEP=-\left( W_{sep}-W_{sep}^{0}\right)=W_{sep}^{0}+\left\langle \dfrac{\mathcal{A}}{N_{gb}}\right\rangle ^{gb}-\left\langle \dfrac{\mathcal{A}}{N_{gb}}\right\rangle ^{fs}.\label{eq:GEP}   
\end{equation}
The GEP is expressed as a per GB site quantity and the minus sign is to conserve the convention that a positive $GEP$ translates into GB embrittlement.

For scenario \ref{fig:fracture}b, concentration is conserved because the fracture event is much faster than the solute diffusion time scale. Also, because the final state is not at equilibrium, we assume that the chemical potential is mainly fixed by bulk atoms and thus unchanged during fracture. The contribution of bulk atoms and solutes to the overall energy of the system is also identical before and after fracture because solutes do not have time to diffuse. The probability of finding a segregated solute at a free surface is identical to the probability of finding a segregated solute at the GB, only the segregation energy changes. 
\begin{equation}
    GEP_b=\sum_{i\in GB}\left(E_{i,fs}^{b}-E_{i,gb}^{b}\right)g_i^{gb}\exp\left(\beta\mathcal{A}_{i,gb}^{b}\right), \label{GEP_b2}
\end{equation}
where the sum runs over excited states $i$ involving GB sites only.

Scenario \ref{fig:fracture}f is a bit more complicated because upon fracture the system initially containing $N_b$ bulk sites, $N_{gb}$ GB sites and $\{ N_{\alpha}\}$ solutes is split into two independent sub-systems, each containing $(N_b-N_{gb})/2$ bulk sites, $N_{gb}$ free surface site and $\{ N_{\alpha}/2\}$ solutes. The concentration in each sub-system after fracture is:
\begin{equation}
    \left[ \alpha\right]^{fs}=\dfrac{\dfrac{N_{\alpha}}{2}}{\dfrac{N_b-N_{gb}}{2}+N_{gb}}=\dfrac{N_{\alpha}}{N_{gb}+N_b}=\left[ \alpha\right]^{gb}. \label{concf}
\end{equation}
Still there is a difference when we solve the LTE equations because the $N_{\phi}/N_{b}$ ratio is not the same in a sub-system containing a free surface and in the initial system containing the GB:
\begin{equation}
\gamma^{fs}=\dfrac{N_{gb}}{\dfrac{N_b-N_{gb}}{2}}=\dfrac{2\gamma^{gb}}{1-\gamma^{gb}}, \label{gammafs}
\end{equation}
with $\gamma^{gb}$ being defined in Eq. \ref{eq:LTE5}. When using Eq. \ref{eq:gi2} to switch from per interface site multiplicities to per total number of sites multiplicities, the value $\gamma^{fs}$ defined in Eq. \ref{gammafs} will be used to compute the grand potential of the system containing a free surface. The GEP for scenario \ref{fig:fracture}f requires two separate LTE calculations (where the same total species concentrations are imposed) and is written as:
\begin{align}
    & GEP_f = \sum_{\alpha}\left[\alpha\right]\left(\mu_{\alpha}^{fs}-\mu_{\alpha}^{gb}\right) \nonumber \\
    & +\sum_{i\in GB} g_i^{\phi} \left(2E_{i,fs}^{b}\exp\left(\beta\mathcal{A}_{i,fs}^{b}\right)-E_{i,gb}^{b}\exp\left(\beta\mathcal{A}_{i,gb}^{b}\right)\right) \nonumber \\
    & +\sum_{i\in bulk} \dfrac{g_i^{b}E_{i,b}^{b}}{\gamma} \left((1-\gamma)\exp\left(\beta\mathcal{A}_{i,fs}^{b}\right)-\exp\left(\beta\mathcal{A}_{i,gb}^{b}\right)\right) , \label{GEP_f2}
\end{align}

\section{Mean-field derivation of geometrical frustration at the interface\label{app_counter}}

In this Appendix we derive a mean-field version of the geometrical frustration at the interface. This contribution must be taken into account in LTE calculations when one type site or configuration has a concentration per site above a few percent. Due to the effect of grain size on segregation (cf. Sec. \ref{subsec:grainsize}) and the high segregation energies of solutes to interfaces (even more so for free surfaces), interface coverage close to unity often arises. With LTEs, one can take into account chemical short-range order exactly; while it is straightforward to do so for a set of well-identified configurations, it is much more complicated to do the combinatorics for "dissociated" configurations (meaning all configurations that are not well-identified). Therefore, in this study, we simplify the calculation of the geometrical multiplicity of these dissociated configurations using a mean-field approach. Short-range order for well-defined configurations is still taken into account exactly. Let us introduce some handy notations to write the derivation in a more concise manner.
\begin{equation}
X=\exp\left(\beta\mu_{\mathrm{P}}\right),
\label{eq4:1}
\end{equation}
\begin{equation}
Y=\exp\left(\beta\mu_{\mathrm{H}}\right),
\label{eq4:2}
\end{equation}
\begin{equation}
B_{PH}=\sum_{i=3,5,7}g_{i}^{b}\exp\left(\beta F^{b}\left(PH_{i}^{b}\right)\right),
\label{eq4:3}
\end{equation}
\begin{equation}
S_{P}=\exp\left(\beta F_{\phi}^{b}\left(P^{\phi}\right)\right),
\label{eq4:5}
\end{equation}
\begin{equation}
S_{H}=3\exp\left(\beta F_{\phi}^{b}\left(H^{\phi}\right)\right),
\label{eq4:6}
\end{equation}
\begin{equation}
S_{PH}=\sum_{i=1,2,3,5}g_{i}^{\phi}\exp\left(\beta F_{\phi}^{b}\left(PH_{i}^{\phi}\right)\right),
\label{eq4:4}
\end{equation}
where $F^{b}$ and $F^{b}_{\phi}$ are the free energies (including vibrational entropy contributions for H) for bulk and interface configurations, with values taken from Tables \ref{tab:bulkenergies} and \ref{tab:gbenergies}. With these handy notations, let us introduce all these configurations--without introducing any geometrical frustration effect--in the expression of the grand potential of the system (Eq. \ref{eq:LTE2}), the $\gamma$ factors being introduced from Eq. \ref{eq:gi2} to take into account the effect of grain radius. 
\begin{equation}
\mathcal{A}=\mathcal{A}_{0}-k_{B}T\ln\left(1+\dfrac{N}{1+\gamma}\left[X+6Y+XYB_{PH}+\gamma Z_{\phi}\right]\right),
\label{eq4:7}
\end{equation}
for which we define the total concentration of segregated solutes at the interface:
\begin{equation}
Z_{\phi}=XS_{P}+YS_{H}+XYS_{PH}.
\label{eq4:8}
\end{equation}
This is where the mean-field approximation comes in: we assume that each segregated solute--whatever its chemical species and/or short-range order--occupies a number of sites at the interface which are therefore no longer available for other species. The exact calculation would require to distinguish between solute type and local configuration. The advantage of this mean-field approximation is that we can use the expressions that have already been used in the previous appendices to take into account geometrical frustration for a single type of solute, the concentration of the unique type of solute becoming the mean quantity of solutes segregated at the interface, $Z_{\phi}$. Hence, Eq. \ref{eq4:7} becomes:
\begin{align}
\mathcal{A}=\mathcal{A}_{0}-k_{B}T\ln\left(1+\dfrac{N}{1+\gamma}\left[X+6Y+XYB_{PH}\right.\right. \nonumber \\
\left.\left.+\gamma\sum_{n\geq1}\dfrac{(N-1)!}{\left(N-n\right)!n!}Z_{\phi}^{n}\right]\right). \label{eq4:9}
\end{align}
As previously, we use the linked-cluster theorem to get rid of all terms that are non-linear in the number of sites in the system $N$,
\begin{equation}
\mathcal{A}=\mathcal{A}_{0}-\dfrac{k_{B}TN}{1+\gamma}\left[X+6Y+XYB_{PH}+\gamma\ln\left(1+Z_{\phi}\right)\right],
\label{eq4:10}
\end{equation}
and the total P concentration is obtained as follows:
\begin{align}
\left[P\right] & =-\dfrac{1}{k_{B}TN}\dfrac{\partial\mathcal{A}}{\partial\ln\left(X\right)} \nonumber \\
& =\dfrac{X}{1+\gamma}\left[1+YB_{PH}+\gamma\dfrac{S_{P}+YS_{PH}}{1+Z_{\phi}}\right].
\label{eq4:11}
\end{align}
The geometrical frustration appears in the $1/(1+Z_{\phi})$ division, and we verify that it is equal to 1 at low interface coverage (i.e. $Z_{\phi}\ll 1$). Note that bulk configurations are always dilute in the system under study, such that a similar treatment is not required.
Equation \ref{eq4:11} can be recast into a second-order polynomial of variable $X$ (first unknown) where each coefficient is a polynomial function of variable $Y$ (second unknown).
\begin{align}
0 & =X^{2}\left(Y^{2}B_{PH}S_{PH}+Y\left(S_{PH}+S_{P}B_{PH}\right)+S_{P}\right) \nonumber \\
 & +X\left(Y^{2}S_{H}B_{PH}+Y\left(\tilde{\gamma}_{P}S_{PH}+B_{PH}+S_{H}\right)+S_{P}\tilde{\gamma}_{P}+1\right) \nonumber \\
 & -\left[P\right]\left(1+\gamma\right)\left(1+YS_{H}\right).
\label{eq4:12}
\end{align}
A similar equation is obtained for the total H concentration:
\begin{align}
0 & =Y^{2}\left(X^{2}B_{PH}S_{PH}+X\left(6S_{PH}+S_{H}B_{PH}\right)+6S_{H}\right)\nonumber \\
 & +Y\left(X^{2}S_{P}B_{PH}+X\left(\tilde{\gamma}_{H}S_{PH}+B_{PH}+6S_{P}\right)+S_{H}\tilde{\gamma}_{H}+6\right) \nonumber \\
 & -\left[P\right]\left(1+\gamma\right)\left(1+XS_{P}\right),\label{eq4:14}
\end{align}
and Eqs. \ref{eq4:12} and \ref{eq4:14} introduce $\tilde{\gamma}_{\alpha}$ for $\alpha=$H or P:
\begin{equation}
\tilde{\gamma}_{\alpha}=\gamma\left(1-\left[\alpha\right]\right)-\left[\alpha\right].
\label{eq4:13}
\end{equation}
Equations \ref{eq4:12} and \ref{eq4:14} correspond to the system of coupled polynomial equations that we solve to obtain P and H chemical potentials at fixed nominal concentrations, and determine whether the presence of one solute alters the segregation behavior of the other.

\bibliography{BIBLIO}
\end{document}